\documentclass[fleqn,12pt]{wlscirep}

\usepackage{amsbsy,amsmath}

\usepackage{subfigure}
\usepackage{xspace, units}
\usepackage{amssymb, amsmath, amsfonts}
\usepackage{lineno}
\usepackage{float}
\usepackage{graphicx}
\usepackage{float}
\usepackage{color}
\usepackage[percent]{overpic}
\usepackage{setspace}
\usepackage{hhline}
\usepackage{array}
\usepackage{blkarray}

\usepackage{graphicx}   
\usepackage{float}

\usepackage{xspace, units}
\usepackage{float}
\usepackage{xcolor}
\usepackage{mathtools}
\usepackage{xcolor}
\usepackage{times}

\usepackage{amsbsy,amsmath}

\usepackage{subfigure}
\usepackage{xspace, units}
\usepackage{amssymb, amsmath, amsfonts}
\usepackage{lineno}
\usepackage{float}
\usepackage{graphicx}
\usepackage{float}
\usepackage{color}
\usepackage[percent]{overpic}
\usepackage{setspace}
\usepackage{hhline}
\usepackage{array}
\usepackage{blkarray}

\usepackage{graphicx}   
\usepackage{float}

\DeclareMathAlphabet{\mathcal}{OMS}{cmsy}{m}{n}

\usepackage{times}
\usepackage{amsbsy,amsmath}
\usepackage{subfigure}
\usepackage{xspace, units}
\usepackage{amssymb, amsmath, amsfonts}
\usepackage{lineno}
\usepackage{float}
\usepackage{graphicx}
\usepackage{float}
\usepackage{color}
\usepackage[percent]{overpic}
\usepackage{setspace}
\usepackage{hhline}
\usepackage{array}
\usepackage{blkarray}
\usepackage{graphicx}   
\usepackage{float}
\usepackage{xspace, units}
\usepackage{float}
\usepackage{xcolor}
\usepackage{mathtools}
\usepackage{xcolor}

\DeclareMathAlphabet{\mathcal}{OMS}{cmsy}{m}{n}

\title{A template-free approach for waveform extraction of gravitational wave events}

\author[1,$\star$]{A. Akhshi}
\author[2]{H. Alimohammadi}
\author[2]{S. Baghram}
\author[2]{S. Rahvar}
\author[2,3,$\dagger$]{M. Reza Rahimi Tabar}
\author[2]{H. Arfaei}
\affil[1]{Department of Physiology, McGill University, Montreal, QC Canada H3G 1Y6}
\affil[2]{Department of Physics, Sharif University of Technology, Tehran 14588, Iran}
\affil[3]{Institute of Physics and ForWind, Carl von Ossietzky University of Oldenburg, Carl-von-Ossietzky-Stra\ss{}e~9--11, 26111~Oldenburg, Germany}
\affil[$\star$]{amin.akhshi@mail.mcgill.ca}
\affil[$\dagger$]{tabar@uni-oldenburg.de}

\begin{abstract}
We develop a general data-driven and template-free method for the extraction of event waveforms in the presence of background noise. Recent gravitational-wave observations provide one of the significant scientific areas requiring data analysis and waveform extraction capability. We use our method to find the waveforms for the reported events from the first, second, and third LIGO observation runs (O1, O2, and O3). Using the instantaneous frequencies derived by the Hilbert transform of the extracted waveforms, we provide the physical time delays between the arrivals of gravitational waves to the detectors.
\end{abstract}

\begin{document} 

\flushbottom
\maketitle
\thispagestyle{empty}
\frenchspacing

\section*{Introduction}

An important prediction of General Relativity is the existence of Gravitational Waves (GWs), which act as ripples of space-time \cite{Einstein1}. LIGO and VIRGO have observationally confirmed this prediction after a century. GWs detection has initiated a new era of research in gravity and has opened a new window to observe and study the Universe. One of the challenging issues in the GW events' extraction is the low amplitude of the ripples and the presence of significant background noise. Therefore, the techniques of signal detection and extraction of its waveform in presence of background noise become paramount.

The method of extracting signals from a given time series has important applications in many fields of science, ranging from neuroscience to astrophysics \cite{FRIEDRICH201187, Tabar2019, Joachim, W1, W2, W2-2, W3}. We are interested in the application of this approach for the detection of GWs \cite{Einstein1, Einstein2, weber, HT, Abram, TW, 2016ApJ818L22A, 2016CQGra33p5004C, 170817, 2017ApJ848L12A, 2016PhRvL116x1102A,abbott2016observing,abbott2016gw150914}. 

To this end, since the first direct detection of GWs by the Laser Interferometer Gravitational-wave Observatory (LIGO) on 14 September 2015, more than fifty-two
``confirmed events"  (from the first, second, and third observation runs O1, O2, O3a, and O3b ) have also been detected  \cite{151226, 170104, 2017ApJ851L35A, 2018arXiv181112907T, 170814, 2018arXiv181112907T, O3-1, O3-2, O3-3, O3-4, abbott2021gwtc}
In this paper, we have  analyzed fifteen publicly available GW events in GWTC-1 and GWTC-2 catalogues from O1, O2, and O3a runs (see Table I).


The experimental efforts to detect GWs go back to Weber's cylindrical bar detector in the 1960s \cite{weber}. However, the efforts were not fruitful due to the weak effect of GWs on the vibration of Weber bar. On the other hand, the indirect evidence for the existence of GWs was provided in the astronomical data concerning the energy loss and hence the change in the rotational frequency of the Hulse-Taylor binary pulsar \cite{HT, TW}. The energy loss matched well with the prediction of general relativity due to the gravitational waves' emission. Thanks to the great progress in the conceptual design of suspended mirrors and optomechanical technology in the Michelson-Morley type experiment of LIGO \cite{Abram}, the direct detection of GW signals has become possible. 

The GWs detection opens new observational windows involving astrophysical events \cite{2016ApJ818L22A} to the large scale structures of the Universe \cite{allen1997detection,2007arXiv0711.1115J, 2008PhRvD..78l4020L, jeong2012large, schmidt2012large, 2014PhRvD..89h2001A, PhysRevD.97.123527, abbott2017upper,2020PhRvD.102l4038C}. Moreover, GWs can place constraints on modified gravity theories \cite{2016CQGra33p5004C, 2016PhRvL116v1101A,barack2019black}. Also, the detection of the GWs from merging binary neutron stars \cite{170817} opens a new area of multi-messenger astrophysics \cite{2017ApJ848L12A}. These observations can also enable us to investigate the classical and quantum properties of Black Holes (BH) \cite{2016PhRvL116x1102A, Abedi:2016hgu}. There is even a possibility that GW candidates observed by LIGO could be due to the merger of primordial black holes which are candidates for dark matter \cite{2016PhRvL116t1301B,2016PhRvL117f1101S} in the possible mass range of $20 M_{\odot}\leq M_{pbh} \leq 100 M_{\odot}$ \cite{2016PhRvL116t1301B,2016PhRvL117f1101S}.

The basis of GW-astronomy's diverse applications is detection, analysis, and interpretation of the GW data. There are several template-dependent and template-free methods, which have been developed in recent years for the detection of gravitational waves \cite{Klimenko_2008, L1, PhysRevD95104046, PhysRevD93122004, 2018JCAP03007C, 2017JCAP08013C, 2018JCAP02013L, 2019arXiv190401683R}. In this work, we develop a data-driven and template-free approach to extract event waveforms from gravitational-wave strain time series available from LIGO and Virgo detectors. 
Furthermore, we use the Hilbert-Huang transform and Hilbert spectrum to extract the time series's instantaneous frequency. The Hilbert spectrum enables us to find the physical time delays of the events in the GW detectors.



\section*{A template-free approach}
\label{Sec:datadriven}
The direct detection based on laser interferometry by LIGO relies on {\it amplitudes} and {\it  phases} of the strain data obtained from two identical setups at Hanford and Livingston (denoted henceforth by ``H" and ``L", respectively) and Virgo detectors (``V") with a standard \textit{template-dependent} analysis \cite{L1}. It is based on the correlation of a set of limited GW template bank and the empirical data, utilizing the optimally matched filter method \cite{2012PhRvD85l2006A}. From matching with templates, one obtains the source properties, and from the time delay between the arrival of the gravitational wave to the detectors, one obtains limited information on the events' angular position, which has led to the emission of the GW.


Our event-waveform detection approach is based on the statistical comparison in the time-frequency domain of background noise and the portion of data that includes the event (Methods). The method consists two pre-processing steps and a spectral subtraction noise reduction algorithm that performs well for extracting any non-noise features from a noisy time series.
 The steps are: (i) First, we use a high-pass filter to filter out frequencies below $30 \,\mathrm{Hz}$. We also use notch filtering to eliminate certain high amplitude spectral lines, potentially disrupting the attempt to search for GW event-waveforms. These narrow resonances (high amplitude spectral lines) are caused by different sources, including harmonics of electrical power, violin modes due to mechanical resonances of the mirror suspensions, and/or spectral lines produced by calibration \cite{abbott2020guide}. (ii) We whiten the raw data, which is equivalent to flattening the spectrum of a given signal, allowing all the bands to participate equally in the power spectrum of a given time series. We applied whitening over time windows of length $8 sec$. This helped the frequency content in data appears equally, and even the smallest contributions could be observed. Time windows of length $8 sec$ was selected because it was relatively large enough compared to GW events duration to avoid undesired boundary effects. While it was comparatively small enough to increase the weight of all frequency bands, including the frequency range of GW events, to have equal weight in the power spectrum \cite{abbott2020guide}. (iii) Finally, we employ a generalized template-independent method to suppress the background noise, which is a combination of a two-step decision-directed noise reduction method \cite{W2,hrnr,W3} using Wiener filtering as its gain function and a noise estimation method based on recursive averaging algorithms analyzing GW events in the presence of uncertainty \cite{W2-2,IMCRA} (Methods). We refer to the final data as {\it processed data}. (iv) From analyzing of the processed data, we provide values for the physical time delays between the arrival of gravitational waves to the detectors, using {\it instantaneous frequencies} of extracted waveforms, derived from Hilbert spectrum \cite{Hil2,Hilbert,Tabar2019} (Methods).


\section*{Results}

Let us demonstrate our approach's applicability to detect the waveform of an event imposed into the background noise with different signal to noise ratio, using the introduced method. We consider a background time series $n(t)$ (the data measured in the either H- or L- or V- detectors) as the stochastic background noise, where its mean value is subtracted. The next step is to superimpose a GW waveform template to the background noise. We take a typical time-series $y(t)$ of one of the generated GW templates by LIGO (for instance, here we use the template for the event observed on 14 September 2015). For superimposed process to generate simulated data from the two time series of $n(t)$ and $y(t)$, we generate a new set of time series $x_{\alpha}(t)=n(t)+\alpha ~ y(t)$, with $\alpha \in [0,1]$. One can interpret the $\alpha$ factor as the strength of the GW template time series compared to the noisy background.
In processing the simulated data, we apply our three steps introduced in previous section. We introduce $\tilde{x}_{\alpha}(t)$ as the new time series (processed data) obtained from $x_{\alpha}(t)$, as well as $\tilde{y}(t)$ and $\tilde{n}(t)$ obtained from the time series of $y(t)$ and $n(t)$ (with a different segment of the background data), respectively.

\begin{figure}[H]
\includegraphics[width=.5\textwidth]{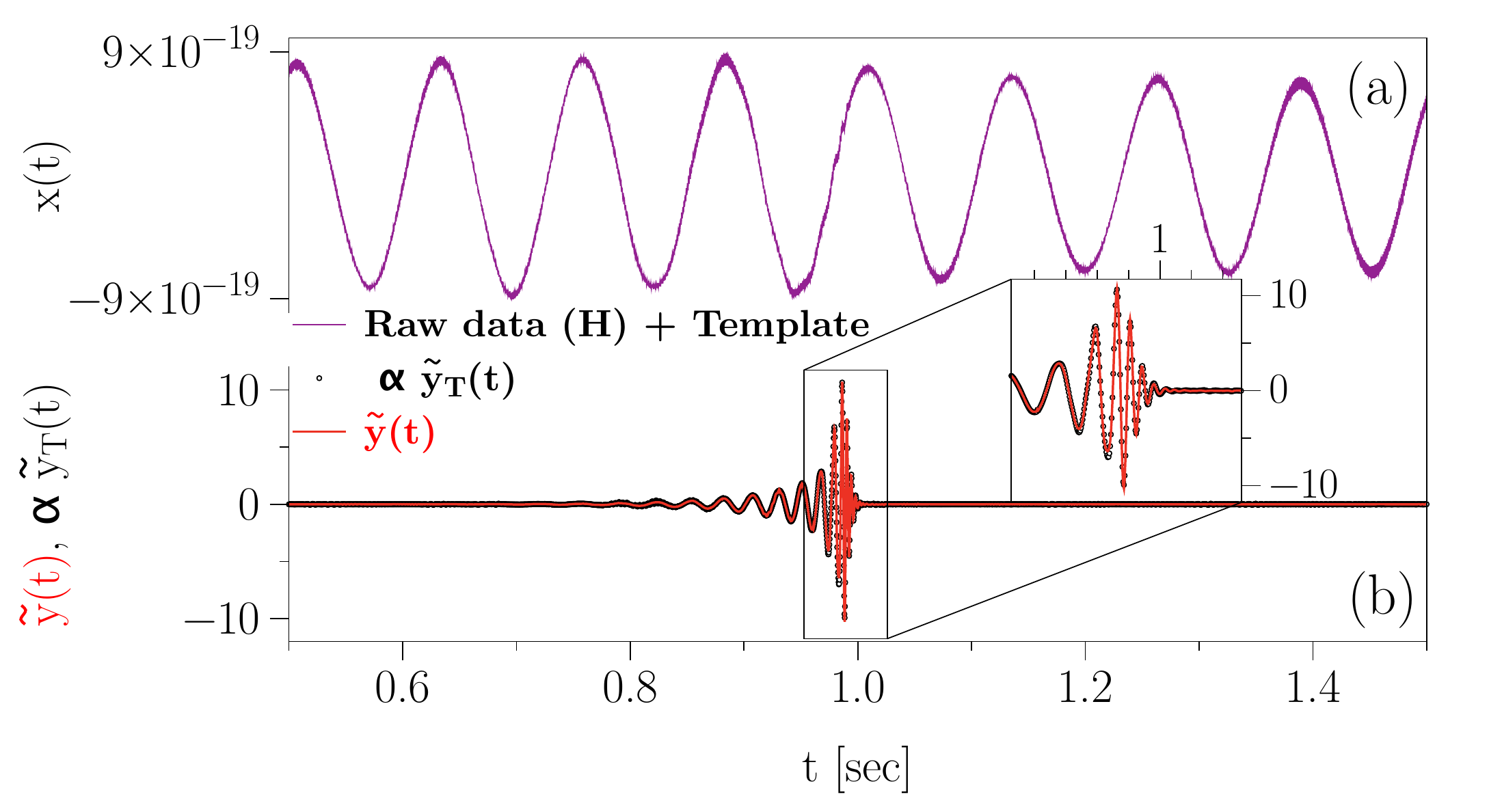}
\includegraphics[width=.5\textwidth]{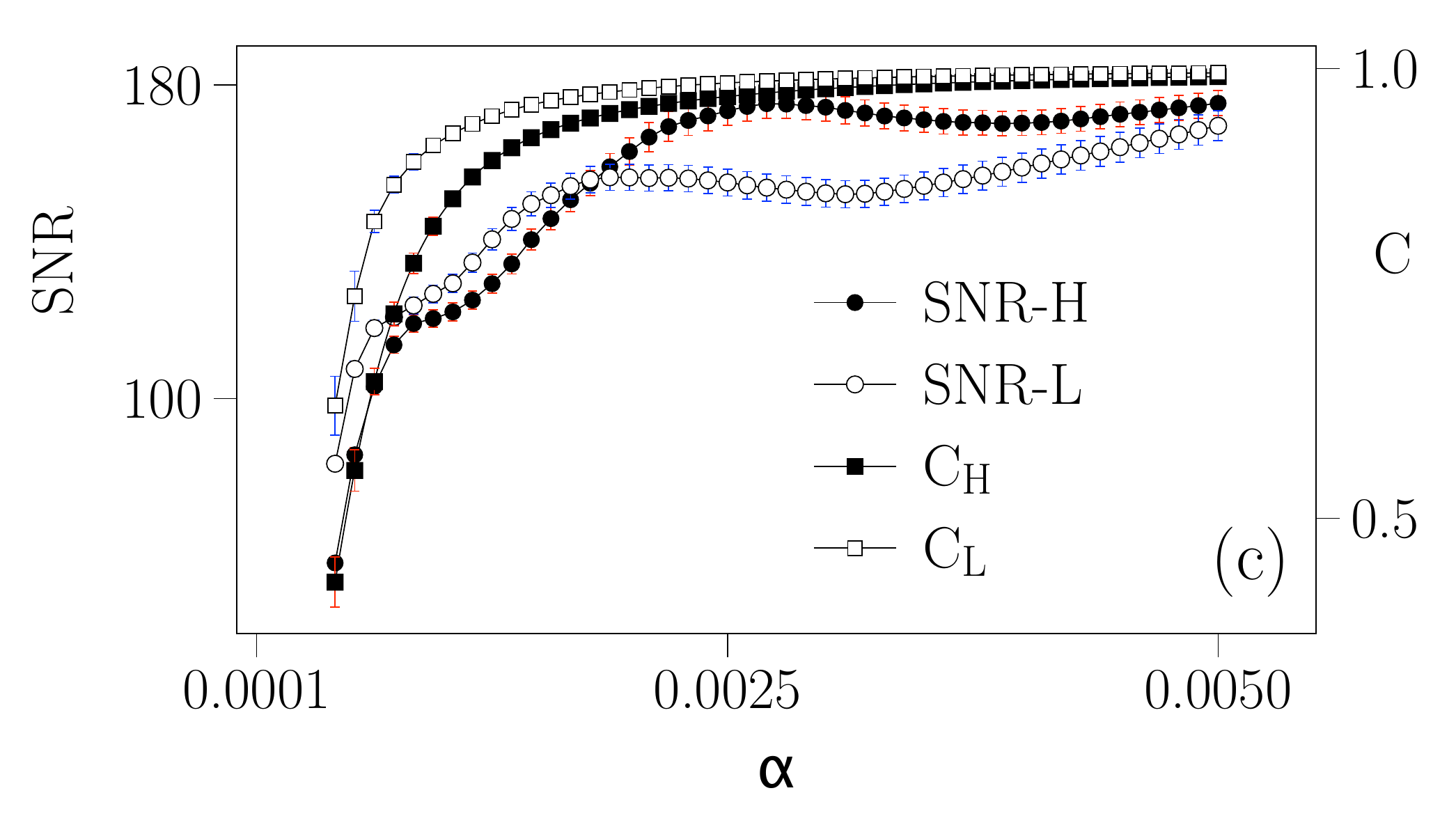}
\caption{ {\bf Extraction of injected signal into background noise with different signal to noise ratio (SNR).}
{\bf (a)} Template of event on 14 September 2015 added to the raw data measured in the H-detector with $\alpha=0.005$. {\bf(b)} Extracted waveform and processed template are also shown. We find a very high similarity between extracted and processed waveforms. The correlation coefficient between two waveforms is $\simeq 0.99$. The extracted waveform is in units of the standard deviation of the processed background noise. 
We applied our proposed steps to extract the injected template out of background noise. We called this an extracted waveform. Also we applied steps of our method only on the simulated template of GW150914 without adding it into background noise (which we called it processed template). {\bf(c)} Signal to noise ratio (SNR) and cross correlation coefficients of extracted waveform with processed template for different $\alpha$s. Error bars indicate the standard error of the mean averaged over 100 ensembles of background noise for each $\alpha$.  We did not preset the time of the injected waveform in 100 ensembles. In this paper SNR is defined as the ratio of extracted waveform and the mean value of processed background noise, $\rho(i,j) = |\hat{y}(i,j)|^2/E[|\hat{n}(i,j)|^2]$, summed over a certain frequency band. 
We note that due to the stochastic behavior of the background noise, different frequency bands in the spectrum will have effectively different SNRs in each time frame. Owing to the nature of our noise estimation approach, the noise will have a nonuniform effect on the SNR ratio of the extracted waveform specially when $\alpha$ is small \cite{W2}. However, as indicated in {\bf(c)} when $\alpha$ increases, this effect will decrease and the relation between the SNR for the extracted waveform and the value of $\alpha$ becomes linear, with
slops about $9300$ and $5300$ for L and H, respectively. }
\label{F1}
\end{figure}

In figure \ref{F1}, we plot the time series $x_{\alpha}(t)$ (in which the template of GW150914, $y(t)$, is imposed), in left upper panel, and extracted waveform, i.e. $\alpha~\tilde{y}_{T}(t)$ denoting the processed waveform, and $\tilde{y}(t)$ denoting the extracted waveform from our method with $\alpha=0.005$ in the left lower panel. The cross-correlation coefficient of extracted waveform with $\tilde{y}(t)$ is about $0.99$ (right panel of figures \ref{F1}). The strains are given in units of the standard deviation of the processed background noise, which provides us the statistical significance of an extracted waveform in each instant. For instance, the maximum amplitude of waveform for $\alpha=0.005$, has about $12$$\sigma$ statistical significance. We note that the one can interpret 
the parameter $\alpha$ as the distance of GW source from the detectors (Methods).

From the spectrum of the processed background noise (i.e. $\tilde{n}(t)$), and the spectrum of the segment of data that includes the event (i.e. $\tilde{y}(t)$), one can define the signal to noise ratio (SNR) $\rho(i,j)$ as $\rho(i,j)= |\tilde{y}(i,j)|^2/E[|\tilde{n}(i,j)|^2]$, where $\tilde{y}(i,j)$ and $\tilde{n}(i,j)$ represent the $j$th spectral component of the time segment $i$ of the processed waveform $\tilde{y}(t)$ and the processed noise $\tilde{n}(t)$ respectively \cite{W2}, and $E[\cdots]$ is the mean operator. The size of time windows are about $0.2$sec and we consider an averaged weighted spectrum $\hat{\sigma^{2}}(i,j)$ over 10 windows with the same size to obtain $E[|\tilde{n}(i,j)|^2]$ (Methods).

We define the integrated signal to noise ratio, $\rho$ in effective frequency bands of 30-500 $Hz$, using thirty equal size-frequency bins \cite{PRX}. This allows us to employ $\chi ^2$ time-frequency discriminator for gravitational wave detection, which enables us to reject the spurious events from the real ones \cite{chi2}. The method applies to each data set and provides the probability $\mathsf{P}$ that the value of $\chi^2$ would be obtained from the chirp signal \cite{chi2}, see Table I and Methods. Our obtained integrated $\rho$ depends on $\alpha$ as shown in \ref{F1} (c). As an example, we simulate the data with GW strains of $\alpha \approx 5 \times10^{-4}$, where the detection algorithm provides the signal to noise ratio $\rho_H \simeq 58$ with the cross-correlation coefficient $\approx 0.99$ between the extracted waveform $\tilde{y}(t)$ and the processed one. The correlation coefficients of extracted waveforms and processed templates for different values of $\alpha$ are given in figure (\ref{F1}) (c). These results demonstrate our approach's high sensibility for the detection of events and the extraction of their waveforms in the time series. The estimated values for $\rho$ using our approach for events reported by LIGO are $\sim 2.4-32.3$, see Table 1.

\begin{figure}[H]
\includegraphics[width=.5\textwidth]{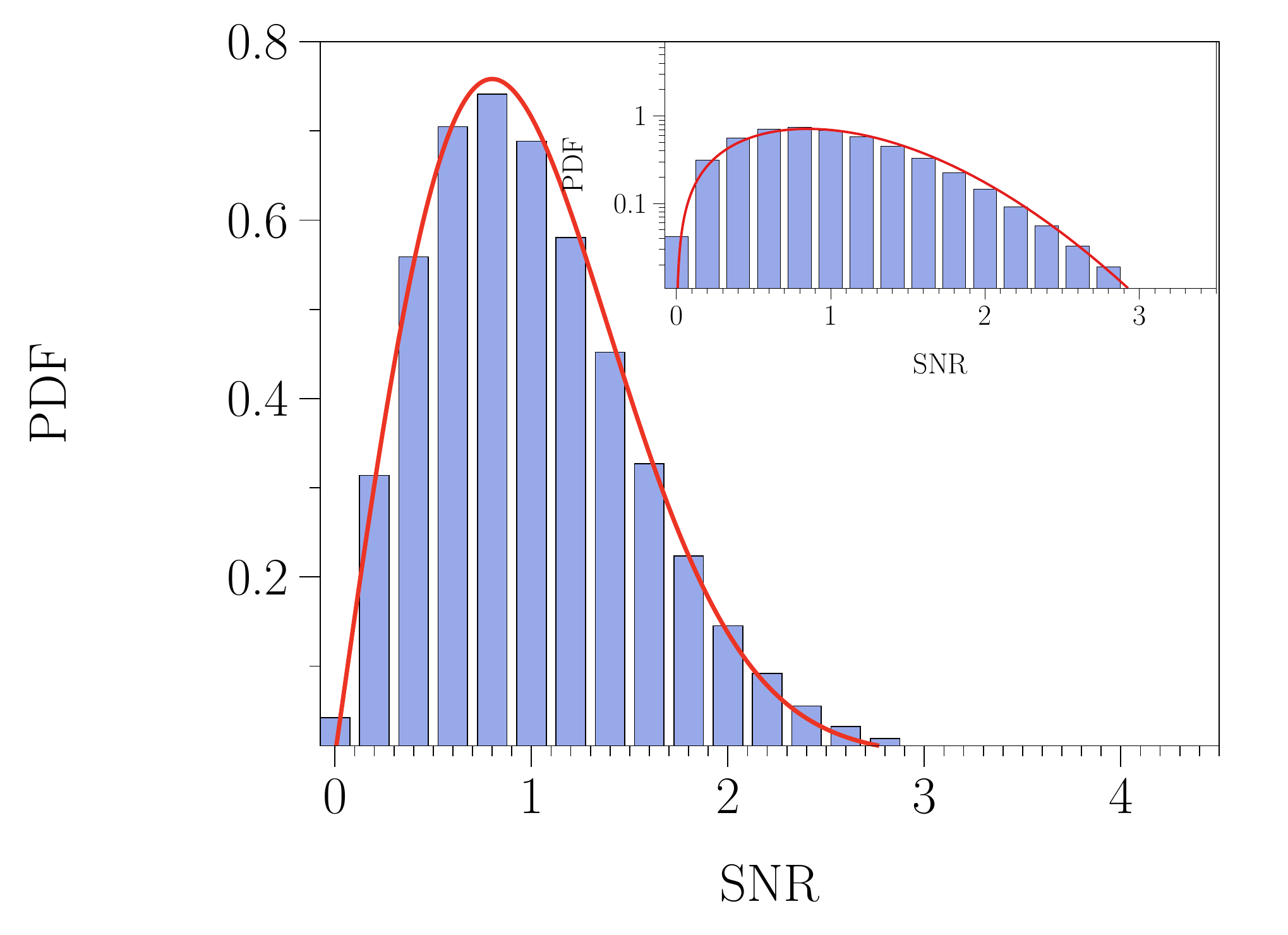}
\caption{ {\bf Probability distribution function (PDF) of signal to noise ratio $\rho$ for processed background noise.}
The PDF of signal to noise ratio $\rho$ for the processed background noise as well as Rayleigh distribution (red) with variance $0.67$. The total time duration of the analyzed time series is about $637$h. The PDF is estimated from $\sim 74\times 10^6$ signal to noise ratio $\rho$ of processed background noise. 
Each SNR is estimated for the 32 frequency bins between $[30 - 500]$Hz for a $10$ seconds segments of time series in O2 and O3a run (see Supp. Info. for the GPS time of analyzed data). Inset the PDF is plotted on a log-linear scale compared to the right-tail of estimated and Rayleigh PDFs. }
\label{F6}
\end{figure}

To reduce the false trigger rate, for a given $\rho^*$, we estimate the joint probability $\mathsf{p}([\rho_H, \rho_L] > \rho^*)$ (i.e. $\rho_H>\rho^*$ and $\rho_L>\rho^*$ ) in shifted time-delay between the detectors, by assuming absence of correlated noise between detectors. This gives the rate of false alert for $\rho^* > 4$ to be $1$ alert per $200$ days of running time \cite{Hanna2008}. The generalization of false alert rate for N-point (N-detector) joint probability $\mathsf{p}$ can be found in \cite{Hanna2008}. By analysing extensive background noise of about  $637$h of data, we find that the $\rho$ has Rayleigh probability distribution function   (see figure \ref{F6} and details are given in Methods).

\begin{figure}[H]
\includegraphics[width=0.9\textwidth]{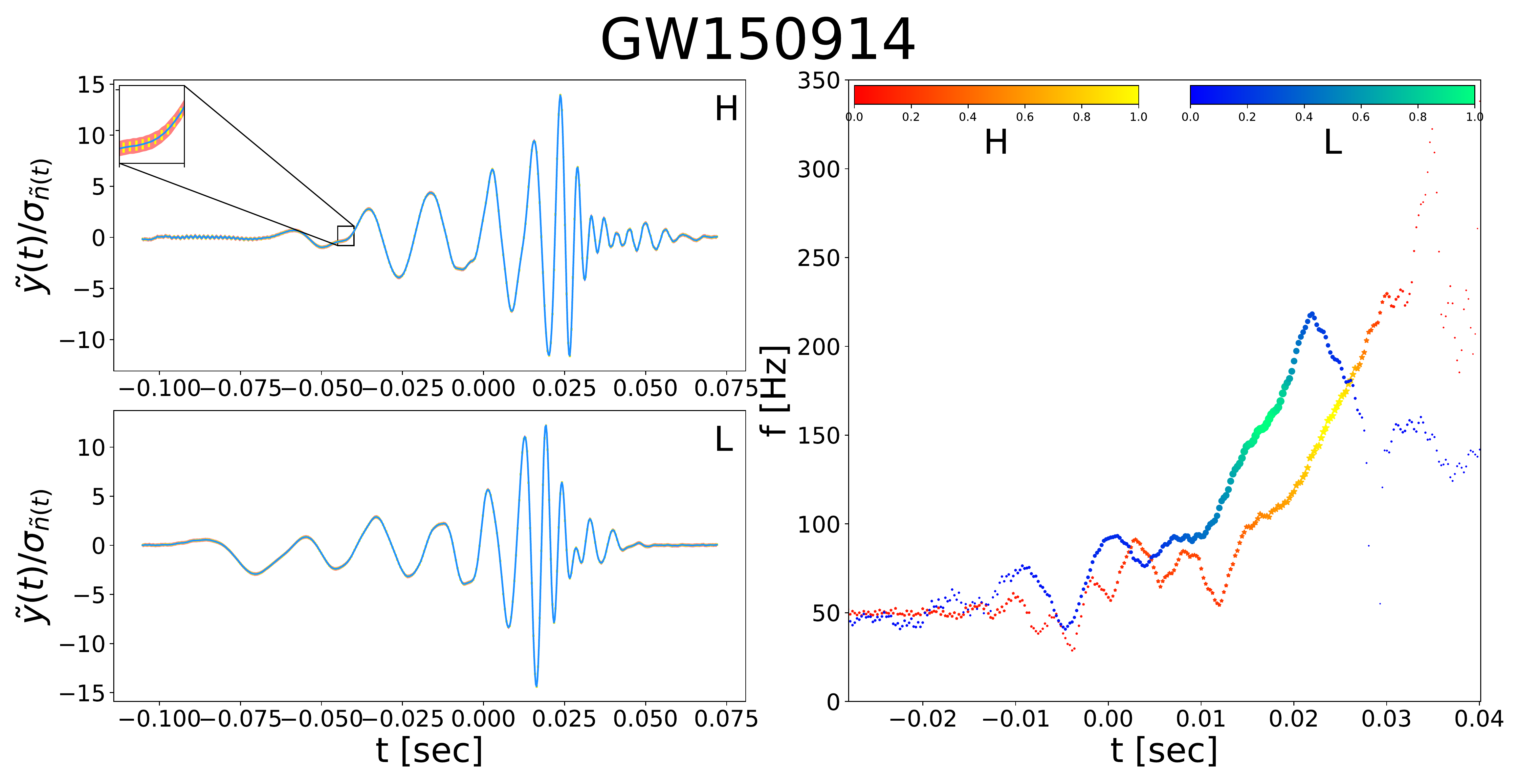}
\caption{ {\bf Extracted waveforms for the event GW150914 in H and L and instantaneous frequency derived by Hilbert spectrum.} Extracted waveforms for the event GW150914 in units of the standard deviation of the processed background noise are given. The $67\%$ (Yellow) and $95\%$ (Orange ) confidence intervals were obtained via bootstrapping. The cross-correlation coefficient between two extracted waveform has value $\simeq -0.92$ in time lag $\simeq 7.3_{-0.5}^{+0.3}\,\mathrm{ms}$ L-first. Cross-correlation coefficients of extracted waveforms in H and L with processed template of event GW150914 are $|C_H| \simeq 0.95$ and $|C_L| \simeq 0.93$, respectively. For extracted waveforms in H and L, instantaneous frequencies derived by the Hilbert spectrum is plotted. Color bar indicating the instantaneous power of the signal at each point in the waveforms. We set the time to zero at the event GPS time reported by LIGO.  
}
\label{F2}
\end{figure}

We apply our method to the GW time series in the LIGO public database. In figure \ref{F2}, we plot the extracted waveforms for GW150914, binary black-hole mergers in H and L detectors. The cross-correlation coefficient between the two extracted waveforms has the value $\simeq -0.92$ with a time-lag $\simeq 7.3_{-0.5}^{+0.3}\,\mathrm{ms}$ where L-detector received the signal first. Also, the cross-correlation coefficients of the extracted waveforms in H and L detectors from our method with the templates of event GW150914, reported by LIGO, are $|C_H| \simeq 0.95$ and $|C_L| \simeq 0.93$, respectively.

\begin{figure}[H]
\includegraphics[width=0.33\textwidth]{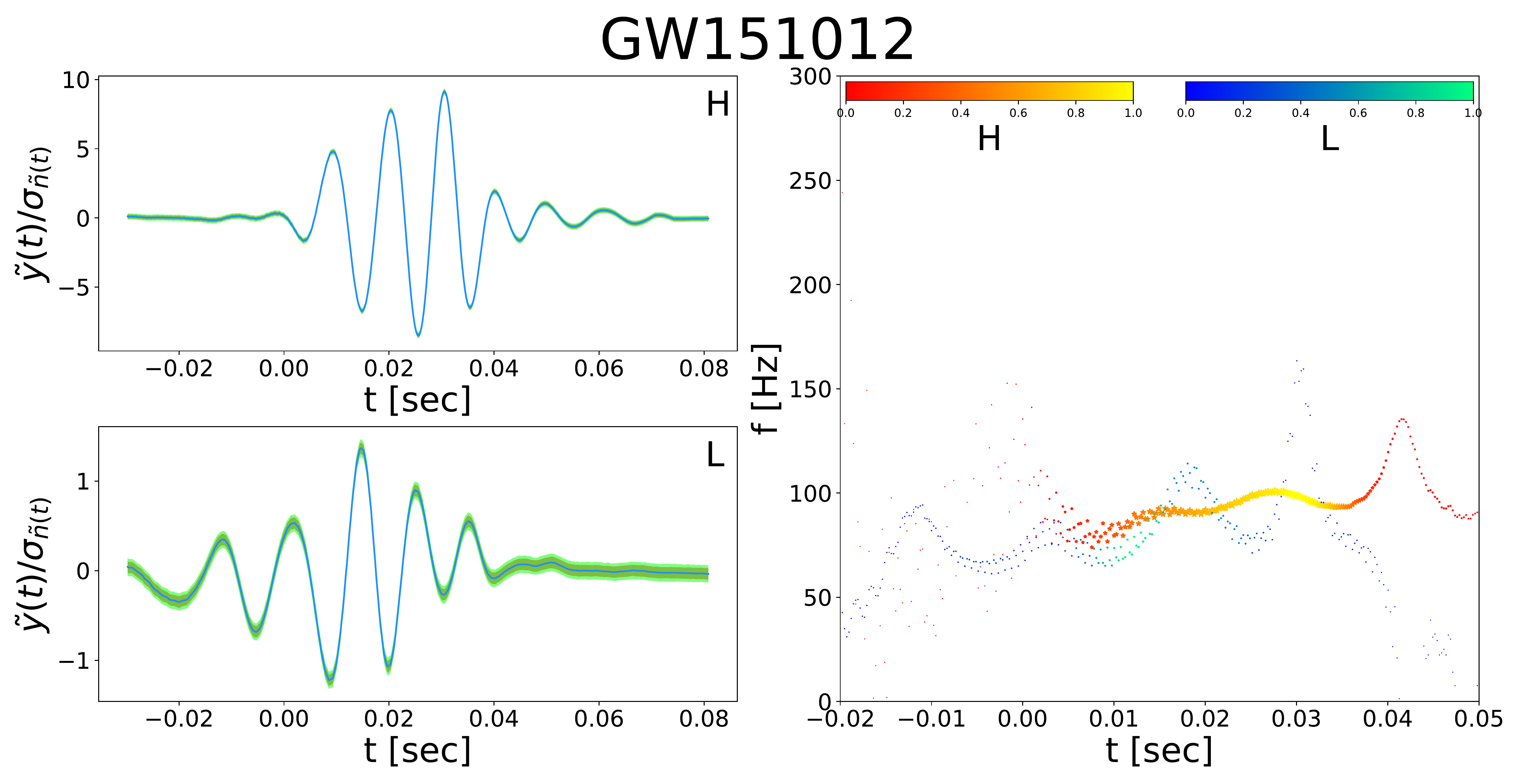}
\includegraphics[width=0.33\textwidth]{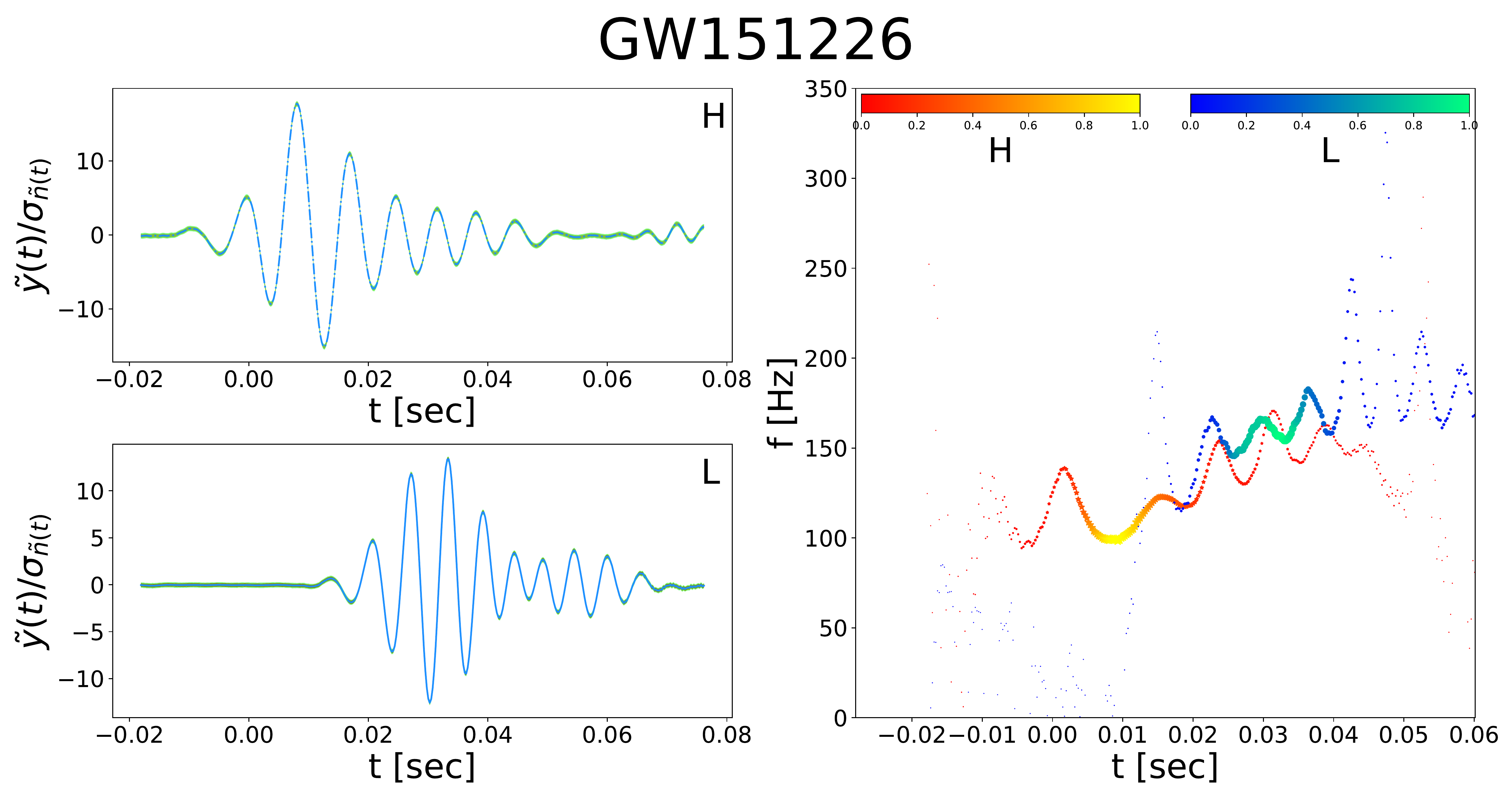}
\includegraphics[width=0.33\textwidth]{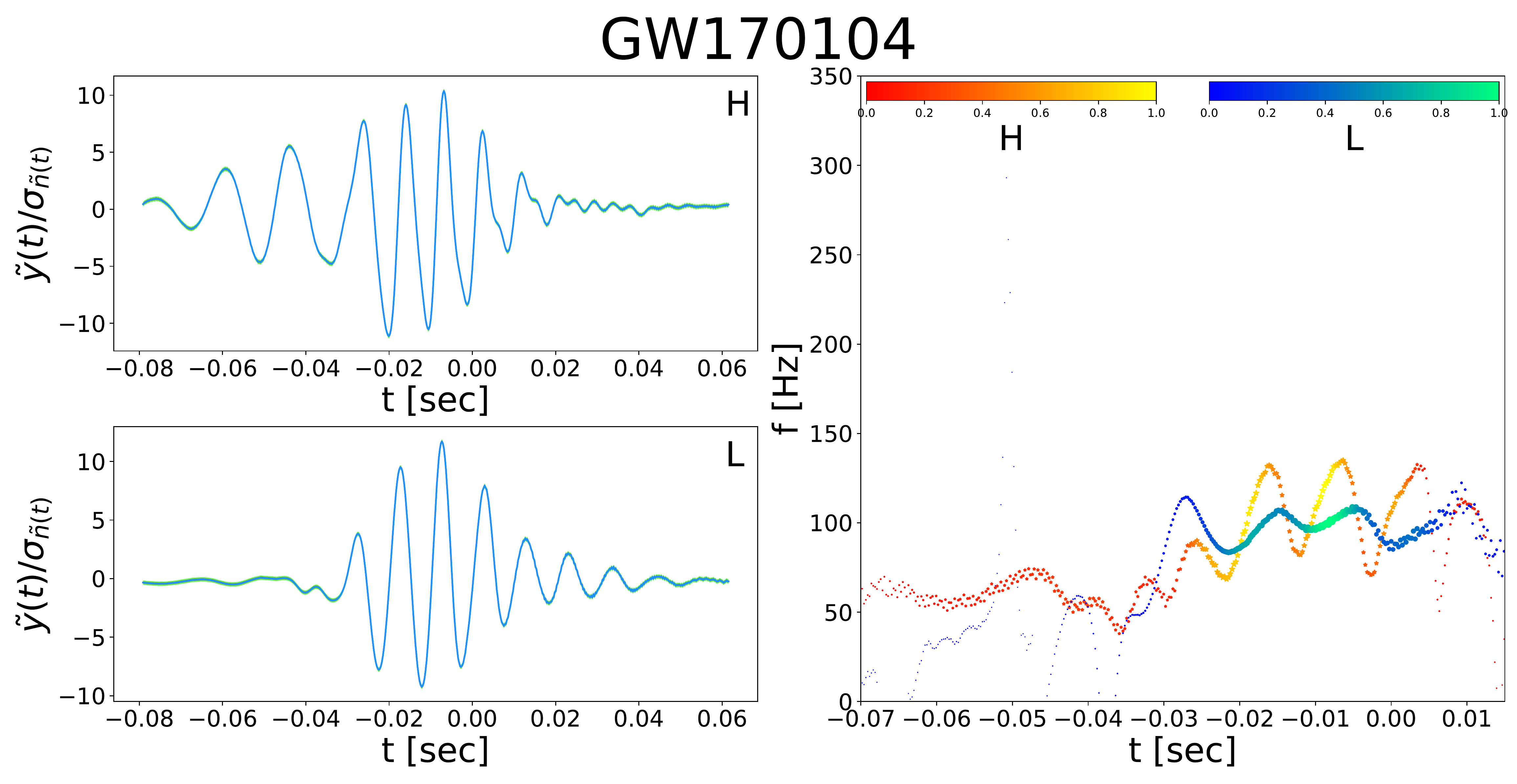}
\includegraphics[width=0.33\textwidth]{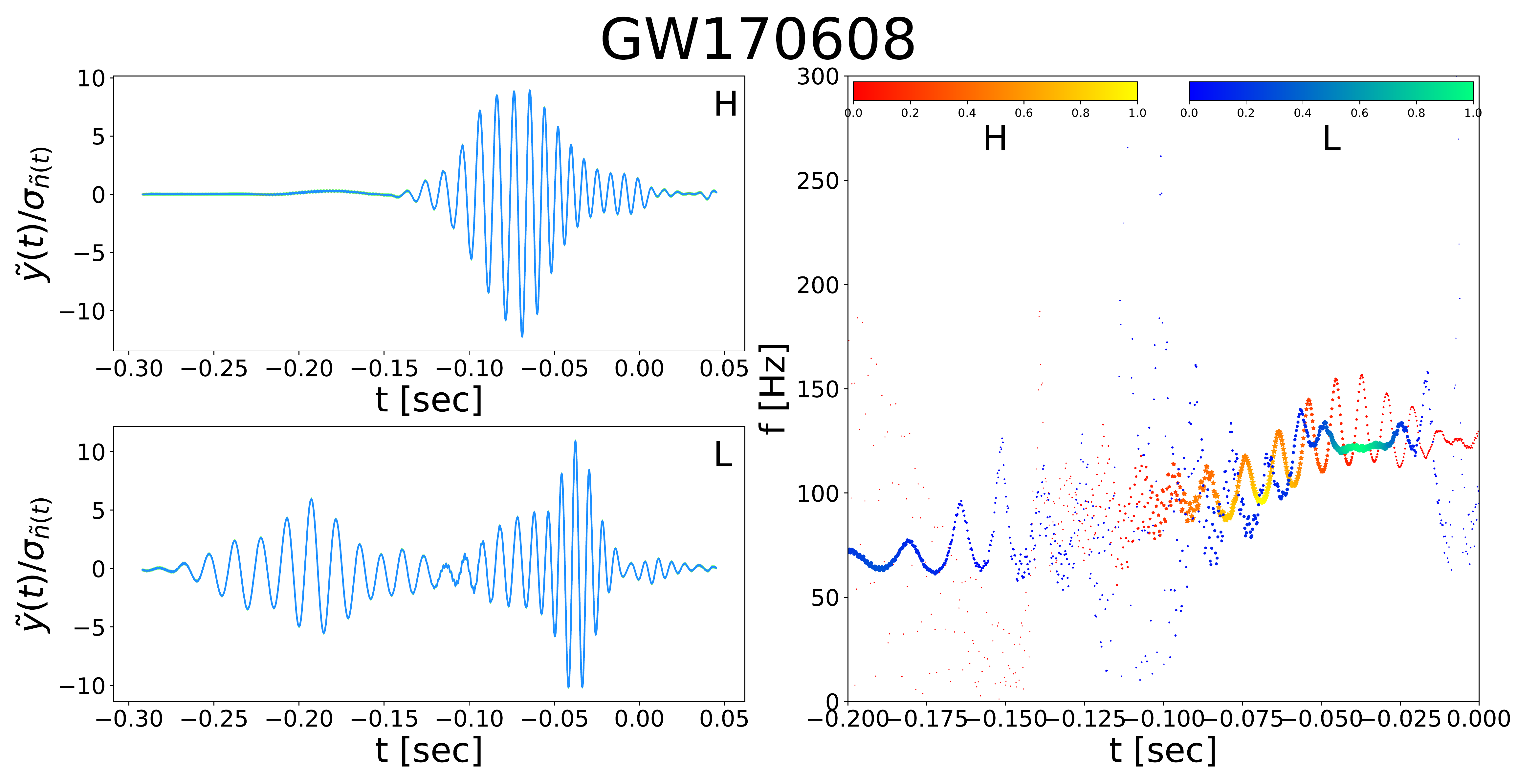}
\includegraphics[width=0.33\textwidth]{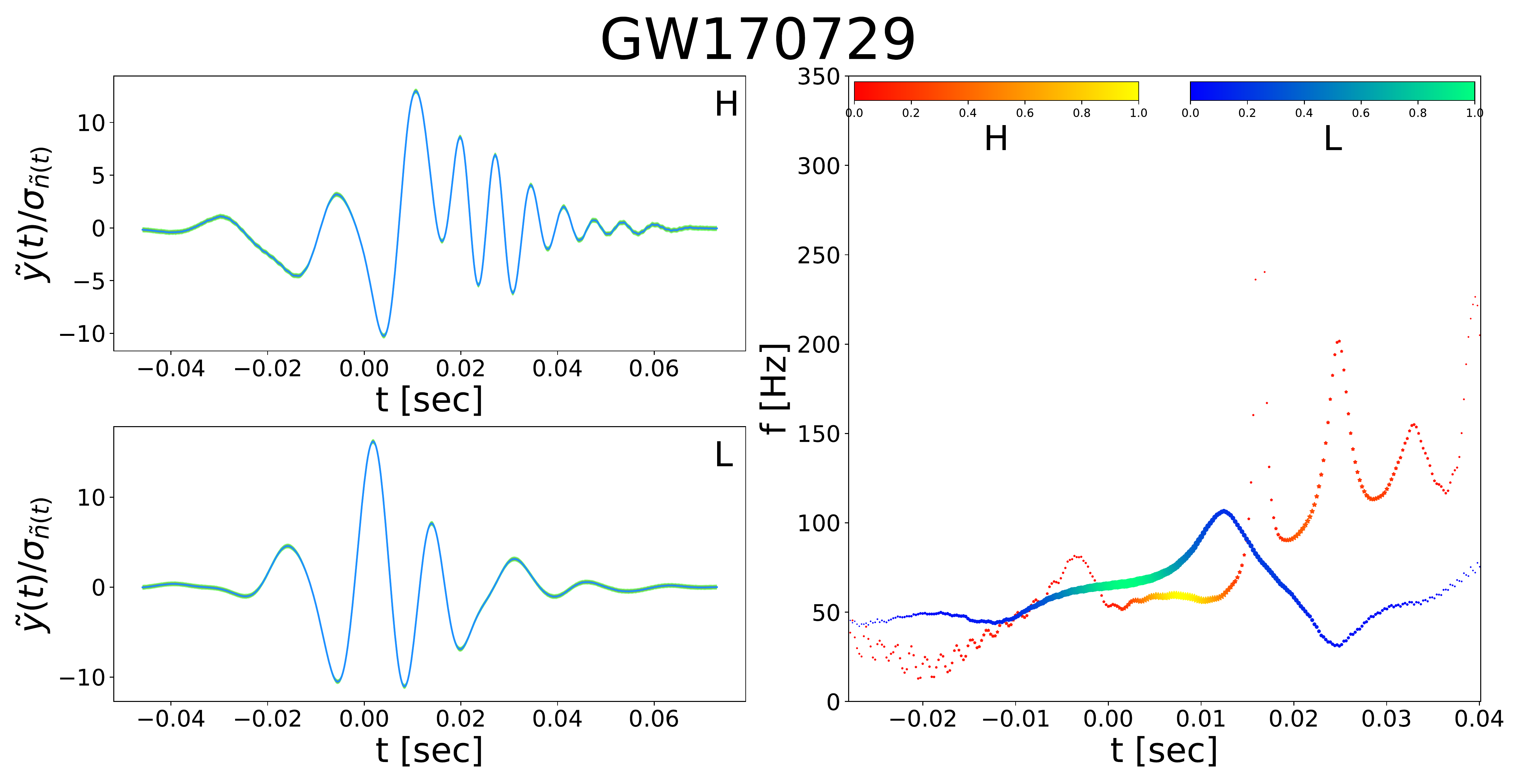}
\includegraphics[width=0.33\textwidth]{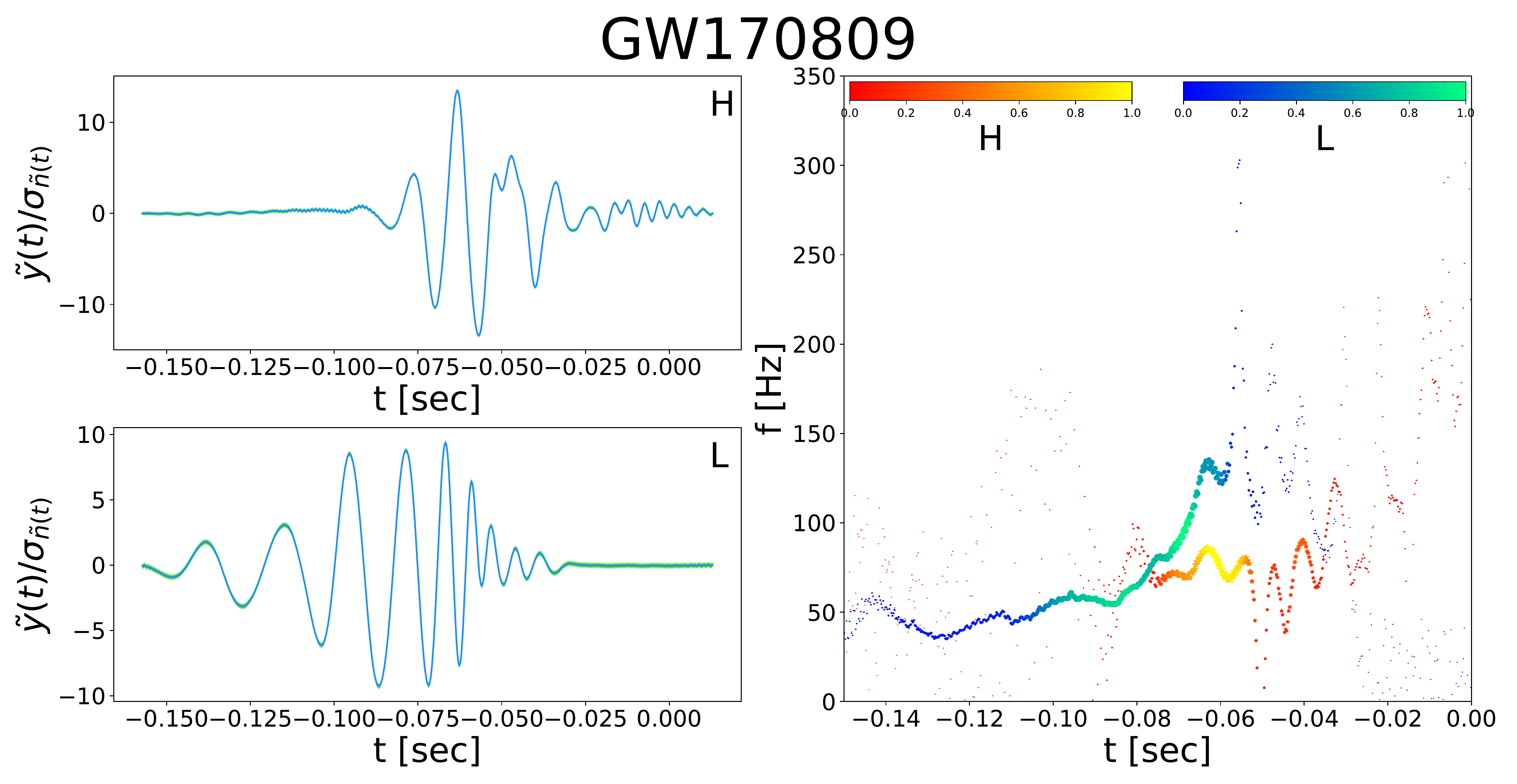}
\includegraphics[width=0.33\textwidth]{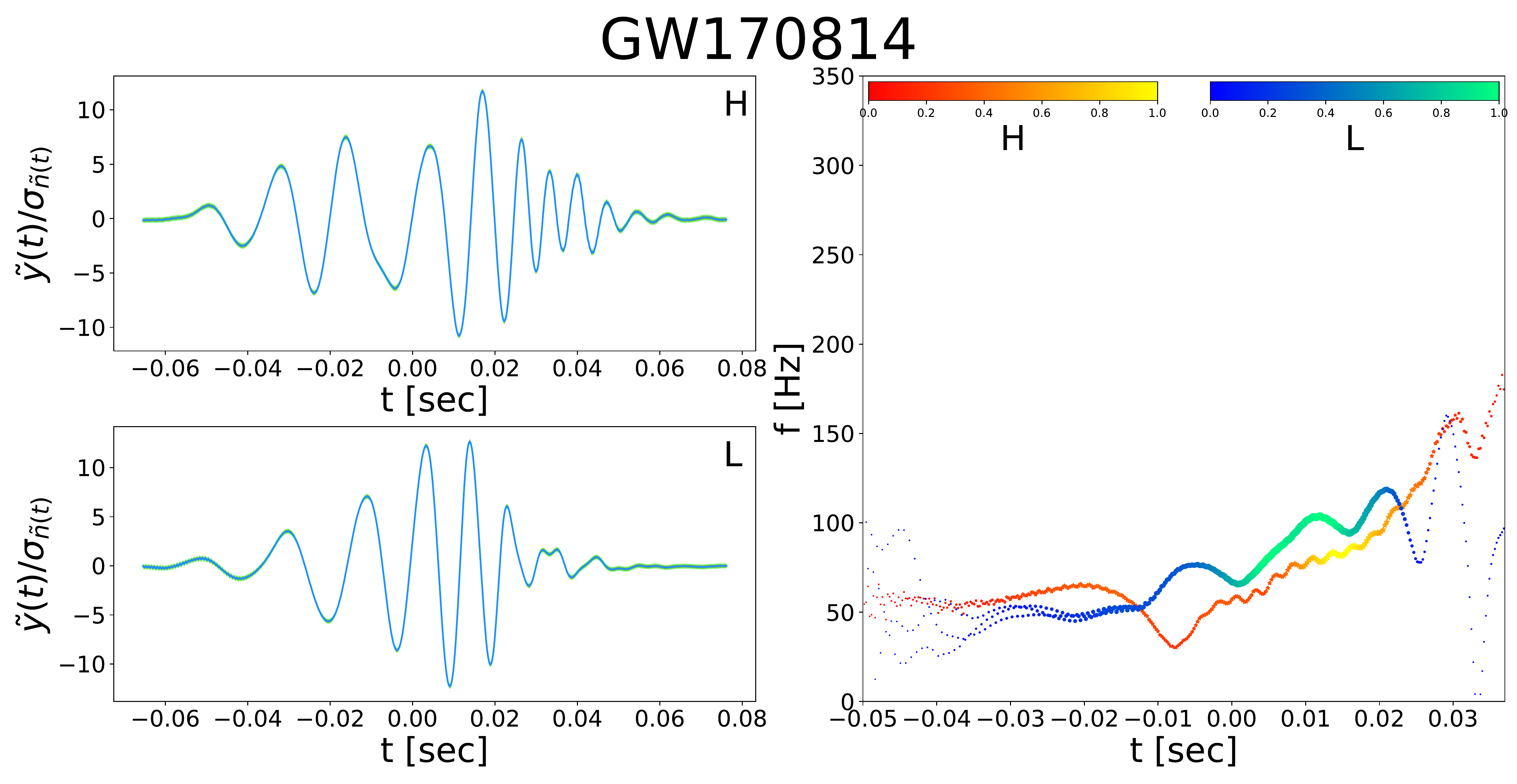}
\includegraphics[width=0.33\textwidth]{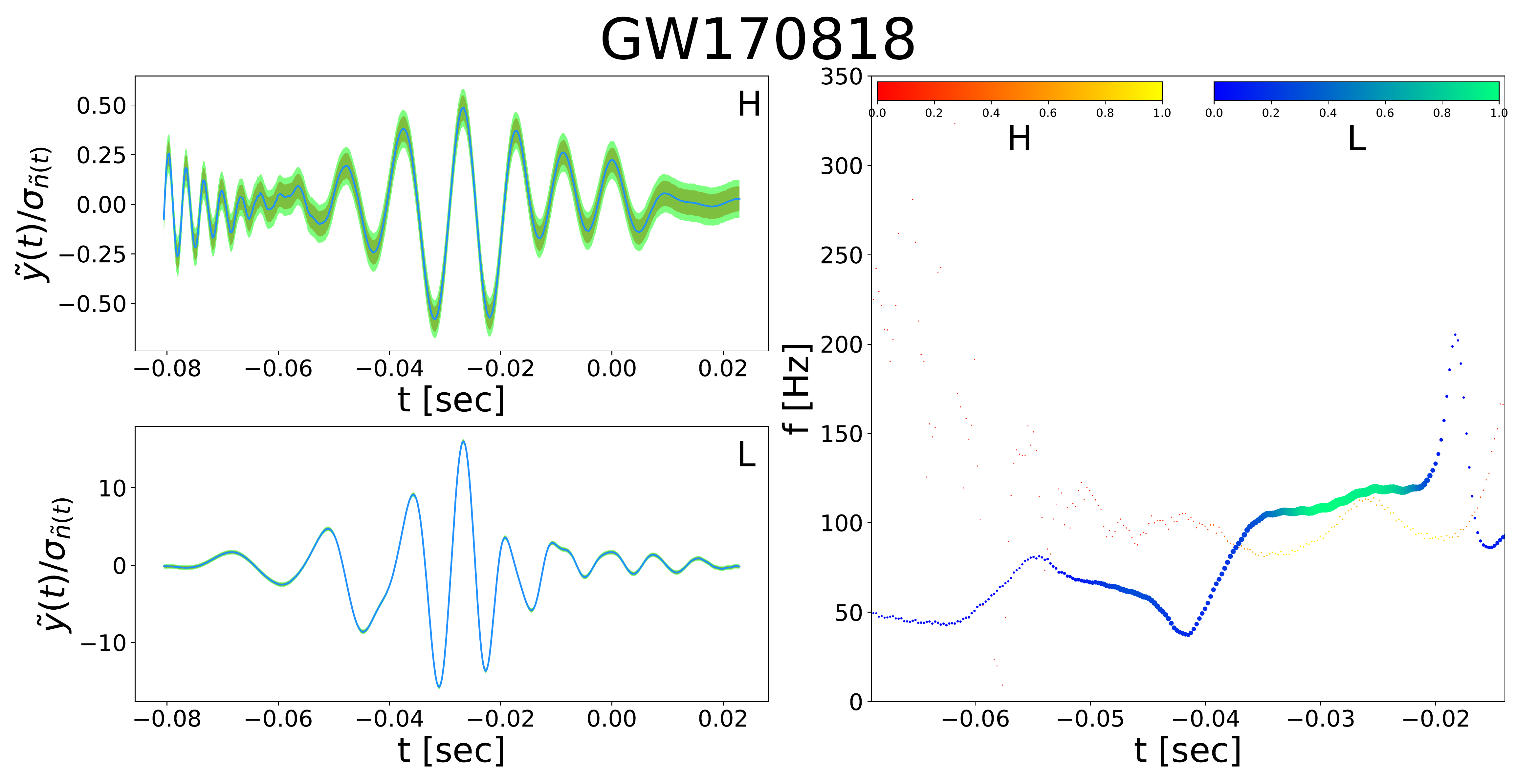}
\includegraphics[width=0.33\textwidth]{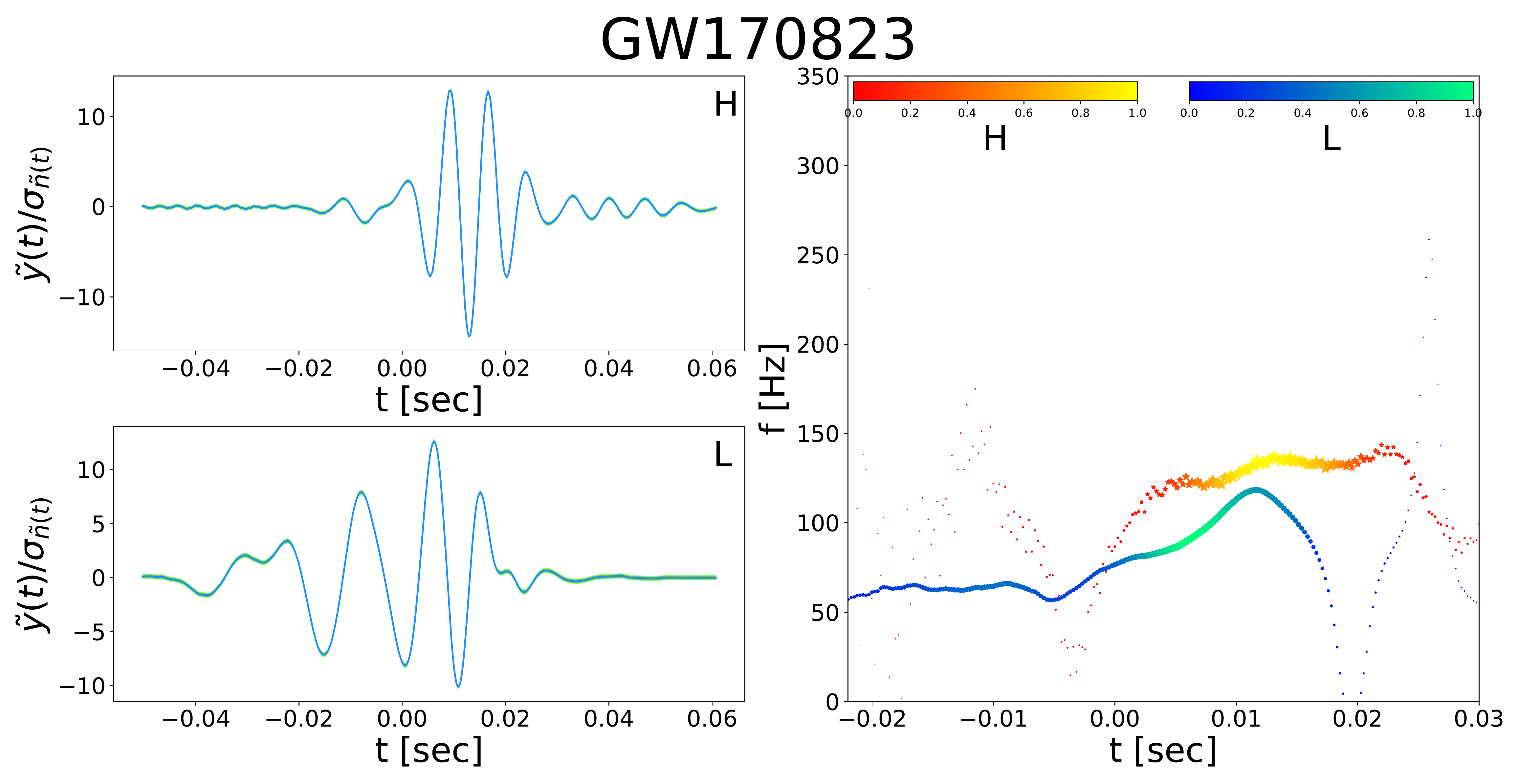}
\includegraphics[width=0.33\textwidth]{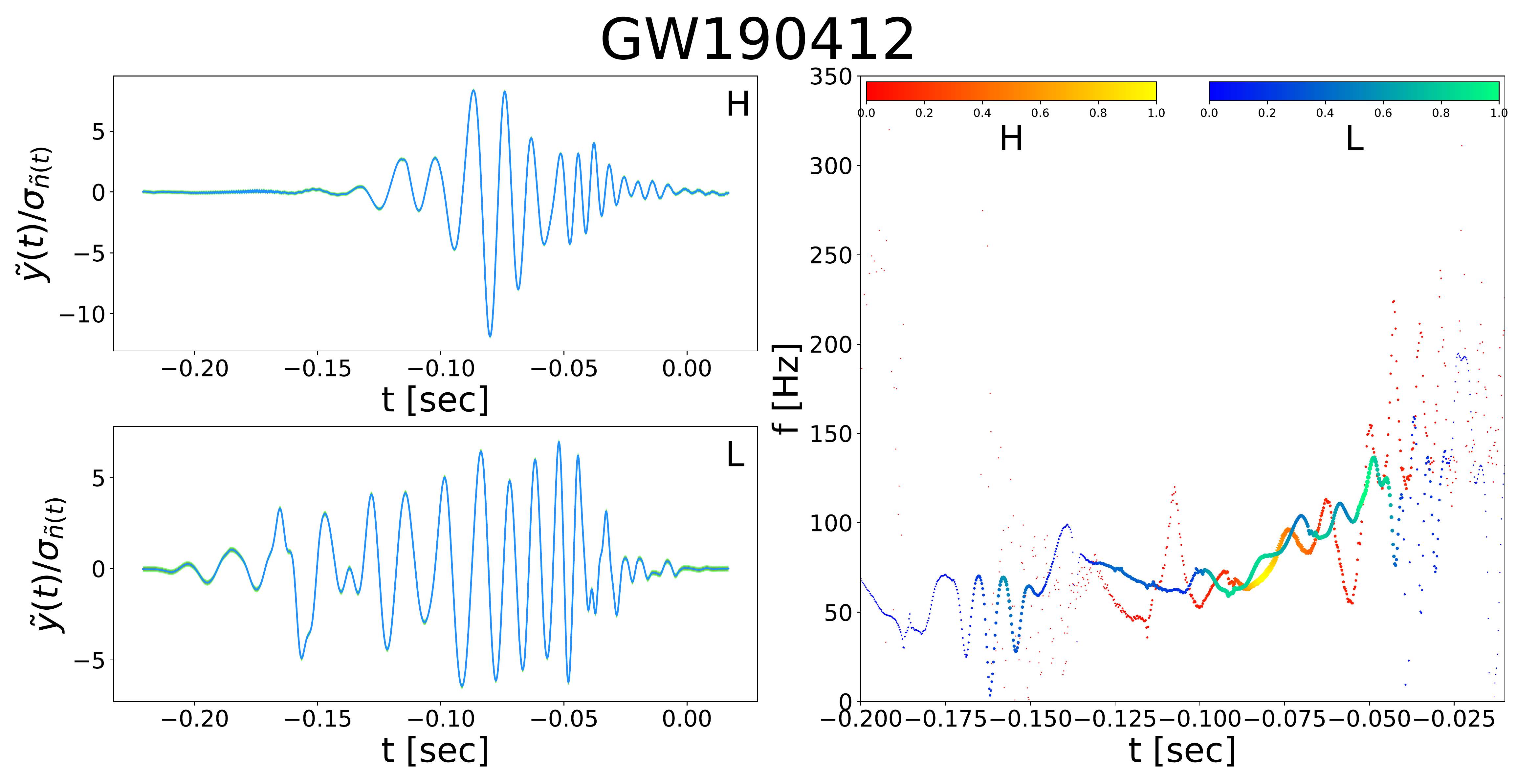}
\includegraphics[width=0.33\textwidth]{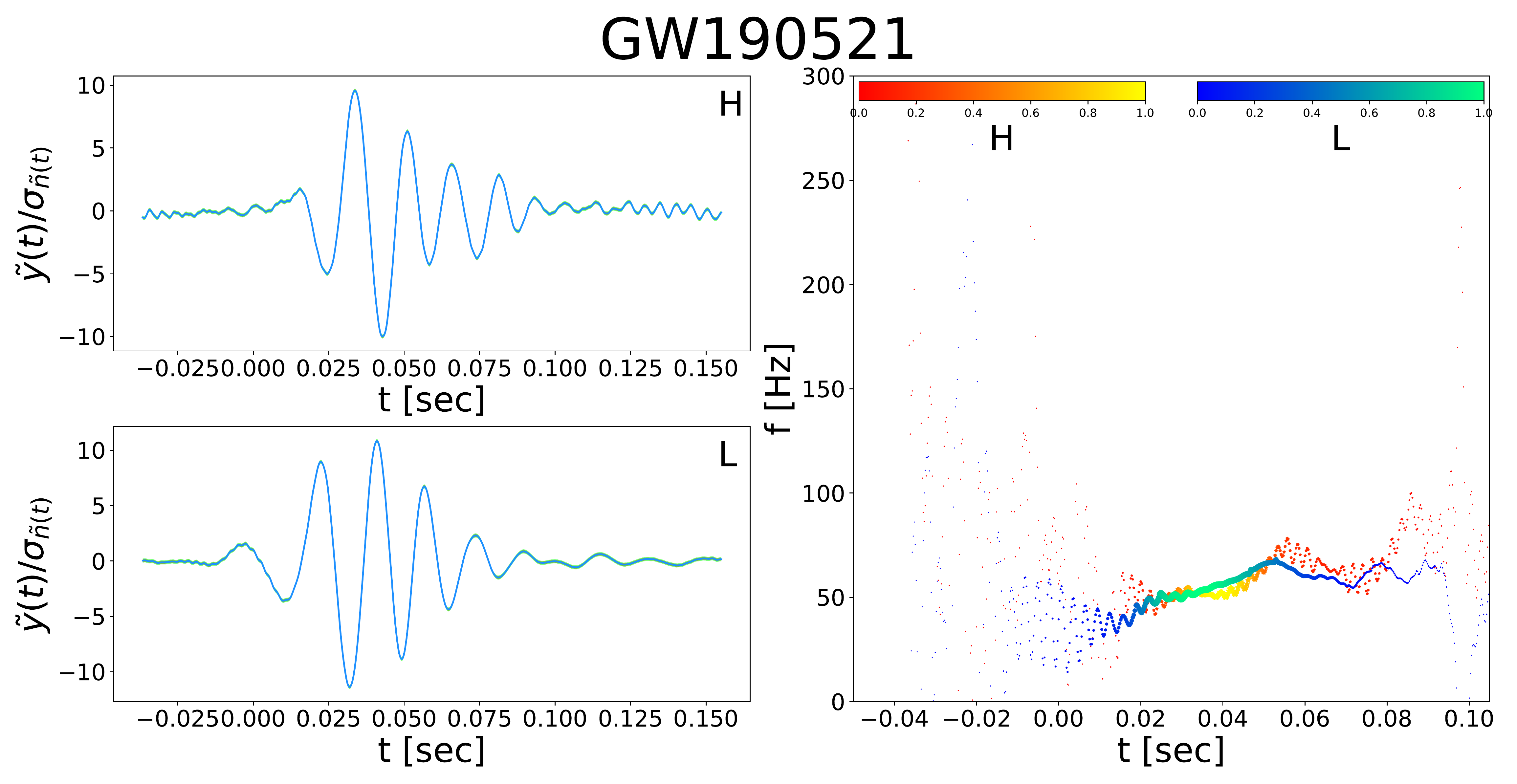}
\includegraphics[width=0.33\textwidth]{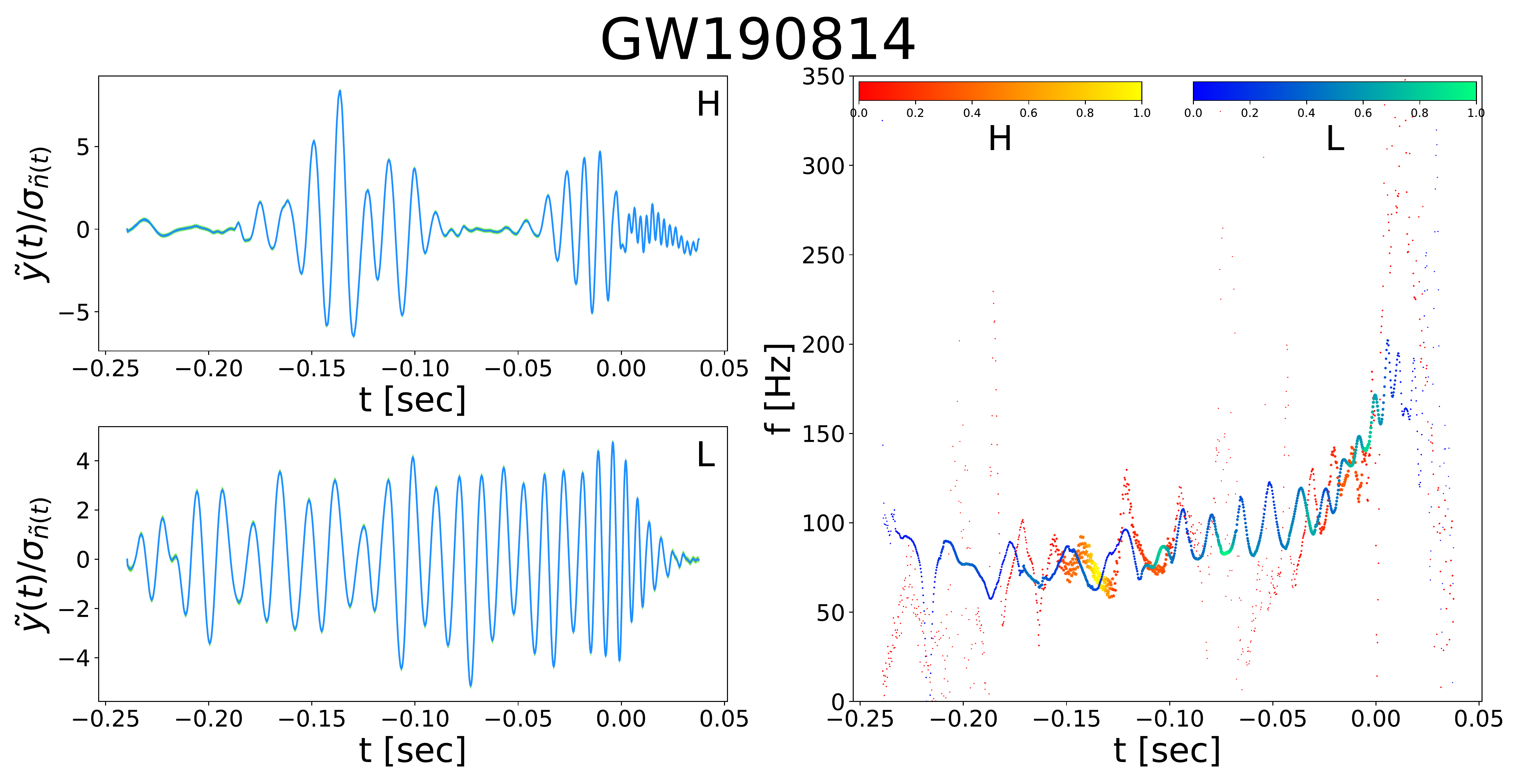}
\caption{ {\bf Extracted waveforms of binary black-hole mergers in observing runs O1, O2 and O3a.}
Extracted waveforms of binary black-hole mergers in O1, O2, and O3a observing runs. They are labeled by the event name and detector that received the signal. The strains are given in units of the standard deviation of the processed background noise. For extracted waveforms in H, L, and V, the Hilbert spectrum's instantaneous frequencies are plotted. The cross-correlation coefficients (R) of extracted waveforms are given in Table 1. We set the time to zero at the event GPS time reported by LIGO.  We excluded waveforms of the GW170817 and the GW190425, since our method was not able to extract clean chirp waveforms for these events.
We are using the information in bivariate time series in which the extracted waveform from one time series is estimated using the information provided by another detector. Therefore in any case that in given time series due to antenna pattern, local noises, etc. it possesses low SNR, this influence the extracted waveform from other time series.
}
\label{F33}
\end{figure}

We perform a similar analysis for the other fourteen events (ten events from the first and second observing runs of O1 and O2 and four events in the third observing run O3a). Our results find thirteen events out of the fifteen reported ones with cross-correlation coefficients between extracted waveforms obtained from our approach and simulated waveforms generated by {\it IMRPhenomD, IMRPhenomPv3HM, IMRPhenomPv2} for GW190814, GW190521, and other remaining GWs, respectively, larger than $0.48$. Their extracted waveforms are depicted in figure \ref{F33}. 
The values of $\mathsf{P}$ and SNR of extracted waveforms are reported in Table I. We excluded waveforms of the GW170817 and the GW190425 from figure \ref{F33}, since our method was not able to extract clean chirp waveforms for these events.



Finally, we determine the time delays between the arrival of the signals to the detectors.
In our method, the physical time delay is estimated from instantaneous frequencies derived from the Hilbert spectrum of the extracted waveforms (see Methods). We determine $f_H(t)$, $f_L(t)$ and $f_V(t)$ and calculate the mean increment in time lags (for instance for H and L) for the condition of $|\tau|< 10\,\mathrm{ms}$, where $C(\tau) = \langle |f_H(t+\tau) - f_L(t)|\rangle$. The value of $\tau$ that minimize (global minimum) $C(\tau)$, is the {\it physical} time-delay. In our analysis, we didn't face the situation that $C(\tau)$ of instantaneous frequencies cross twice or having two minima with the same depths.

In figures \ref{F2}, \ref{F33}, we plot the instantaneous frequency derived from the Hilbert spectrum for all events in the first and second observing runs of O1, O2 and O3a (Methods).

\begingroup
\begin{table*}[!ht]
\caption{Results of our approach for detection of fifteen GW candidates.
 n/a time delays in the LIGO column means the corresponding time delay for that GW event is not reported by LIGO, at least we were not able to find it. n/a in the physical delay column means we didn't find a statistical meaningful time delay to report for that GW event.
}
\resizebox{1\columnwidth}{!}{
\begin{tabular}{l c c c c c}
Event & \multicolumn{2}{c}{Delay ($\mathrm{ms}$)} & P \% & SNR &  R\\
\cline{2-3} & LIGO &  physical delay \\
\hline
{\color{magenta} GW150914} \cite{L1} & $6.9^{+0.5}_{-0.4}$ -- L-first & $ 7.3_{-0.5}^{+0.3}$-- L-first & $\mathsf{P}_H=0.99$, $\mathsf{P}_L=0.79$, & $\rho_H=12.1, \rho_L=14.3$ & $0.92_{-0.007}^{+0.007}$ \\
{\color{magenta} GW151012} \cite{2016PhRvX6d1015A} & $0.6^{+0.6}_{-0.6}$ & $ 0.5_{-0.3}^{+0.5}$ -- L-first & $\mathsf{P}_H=0.78$, $\mathsf{P}_L=0.99$, & $\rho_H=19.6, \rho_L=10.2$ & $0.68_{-0.005}^{+0.005}$ \\
{\color{magenta} GW151226} \cite{151226}& $1.1^{+0.3}_{-0.3}$ -- L-first & $ 1.2_{-0.5}^{+0.7}$ -- L-first & $\mathsf{P}_H=0.99$, $\mathsf{P}_L=0.97$,& $\rho_H=11.9 , \rho_L=10.6$ & $0.33_{-0.008}^{+0.008}$ \\
{\color{magenta} GW170104} \cite{170104}& $3.0^{+0.4}_{-0.5}$ -- H-first & $ 3.2_{-0.2}^{+0.5}$ -- H-first & $\mathsf{P}_H=0.83$, $\mathsf{P}_L=0.81$, & $\rho_H=21.8, \rho_L= 28.5$ & $0.86_{-0.013}^{+0.013}$\\
{\color{magenta} GW170608} \cite{2017ApJ851L35A}& $7$ -- H-first & $ 6.8_{-0.5}^{+0.2}$ -- H-first & $\mathsf{P}_H=0.99$, $\mathsf{P}_L=0.99$, & $\rho_H=5.9 , \rho_L=3.0$ & $0.72_{-0.004}^{+0.004}$\\
{\color{magenta} GW170729} \cite{2018arXiv181112907T}& n/a & $ 1.8_{-0.9}^{+1.0}$ -- L-first & $\mathsf{P}_H=0.99$, $\mathsf{P}_L=0.99$, & $\rho_H=13.7, \rho_L=17.7$ & $0.60_{-0.023}^{+0.023}$\\
{\color{magenta} GW170809} \cite{2018arXiv181112907T} & n/a & $ 9.5_{-0.5}^{+0.5}$ -- L-first & $\mathsf{P}_H=0.99$, $\mathsf{P}_L=0.88$, & $\rho_H=7.9, \rho_L=9.8$ & $0.63_{-0.023}^{+0.023}$\\
{\color{magenta} GW170814} \cite{170814} & $8$ -- L-first & $ 7.8_{-0.5}^{+0.8}$ -- L-first & $\mathsf{P}_H=0.99$, $\mathsf{P}_L=0.99$, &$\rho_H=8.8, \rho_L=5.4$ & $0.76_{-0.011}^{+0.011}$\\
{\color{magenta} GW170817} \cite{170817} & n/a &  n/a & $\mathsf{P}_H=10^{-3}$, $\mathsf{P}_L=0.92$,&$\rho_H=32.3 , \rho_L=2.4$ & $0.01_{-0.10}^{+0.10}$\\
{\color{magenta} GW170818} \cite{2018arXiv181112907T} & n/a & $4.8_{-0.8}^{+.5}$ -- L-first & $\mathsf{P}_H=0.99$, $\mathsf{P}_L=0.70$, & $\rho_H=4.6 , \rho_L=21.0$ & $0.70_{-0.023}^{+0.023}$\\
{\color{magenta} GW170823} \cite{2018arXiv181112907T}& n/a & $ 1.5_{-0.5}^{+0.7}$ -- H-first & $\mathsf{P}_H=0.71$, $\mathsf{P}_L=0.99$, & $\rho_H=21.3 , \rho_L=9.8$ & $0.55_{-0.022}^{+0.022}$\\
{\color{magenta} GW190412} \cite{O3-1}& n/a & $4.0_{-0.2}^{+0.3}$ -- L-first & $\mathsf{P}_H=0.98$, $\mathsf{P}_L=0.93$, & $\rho_H=7.0 , \rho_L=21.8$ & $0.79_{-0.002}^{+0.002}$\\
{\color{magenta} GW190425}\cite{O3-2}& n/a & n/a & $\mathsf{P}_V=10^{-3}$, $\mathsf{P}_L=0.50$, & $\rho_V=32.1 , \rho_L=6.6$ & $0.01_{-0.13}^{+0.13}$\\
{\color{magenta} GW190521} \cite{O3-4}& n/a & $2.0_{-0.4}^{+0.3}$ -- L-first & $\mathsf{P}_H=0.99$, $\mathsf{P}_L=0.99$, & $\rho_H=6.5, \rho_L=4.8$ & $0.95_{-0.002}^{+0.002}$\\
{\color{magenta} GW190814} \cite{O3-3}& n/a & $3.4_{-0.6}^{+0.5}$ -- H-first & $\mathsf{P}_H=0.94$, $\mathsf{P}_L=0.99$, & $\rho_H=14.3, \rho_L=8.4$ & $0.49_{-0.062}^{+0.062}$\\
\end{tabular}
}
\end{table*}\label{tab1}
\endgroup


\section*{Discussion}

\begin{figure}[H]
\centering
\includegraphics[width=.9\textwidth]{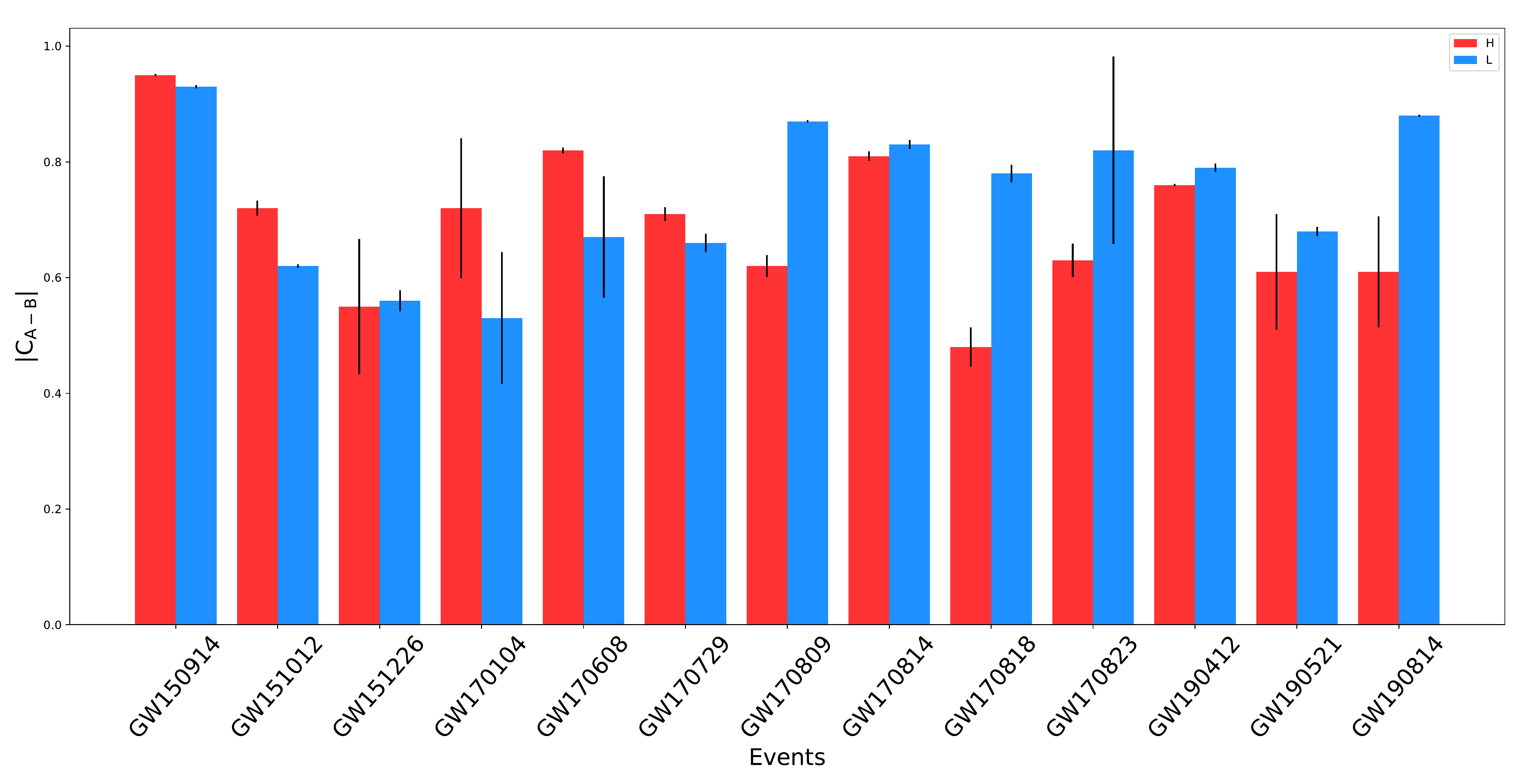}
\caption{ {\bf Cross correlation coefficients of the extracted waveforms for the events in the first, second and third observing runs O1,O2 and O3a by LIGO with waveforms generated by {\it PyCBC} library.}
Cross-correlation coefficients of the extracted waveforms for the event reported by LIGO with simulated waveforms generated by {\it PyCBC} library, where their waveforms are given in figure \ref{F33}. For all events, $A$ refers to the extracted waveform in H or L and $B$ refers to simulated waveform for each detector. \cite{pycbc}
Correlation coefficients between waveforms extracted from different detectors depend on two general parameters. One is the antenna pattern or the spatial angle between the detector and the GW which receives there. The other one is the SNR ratio of the recorded GW in each detector. 
 }
\label{F5}
\end{figure}

\begin{figure}[H]
\centering
\includegraphics[width=.7\textwidth]{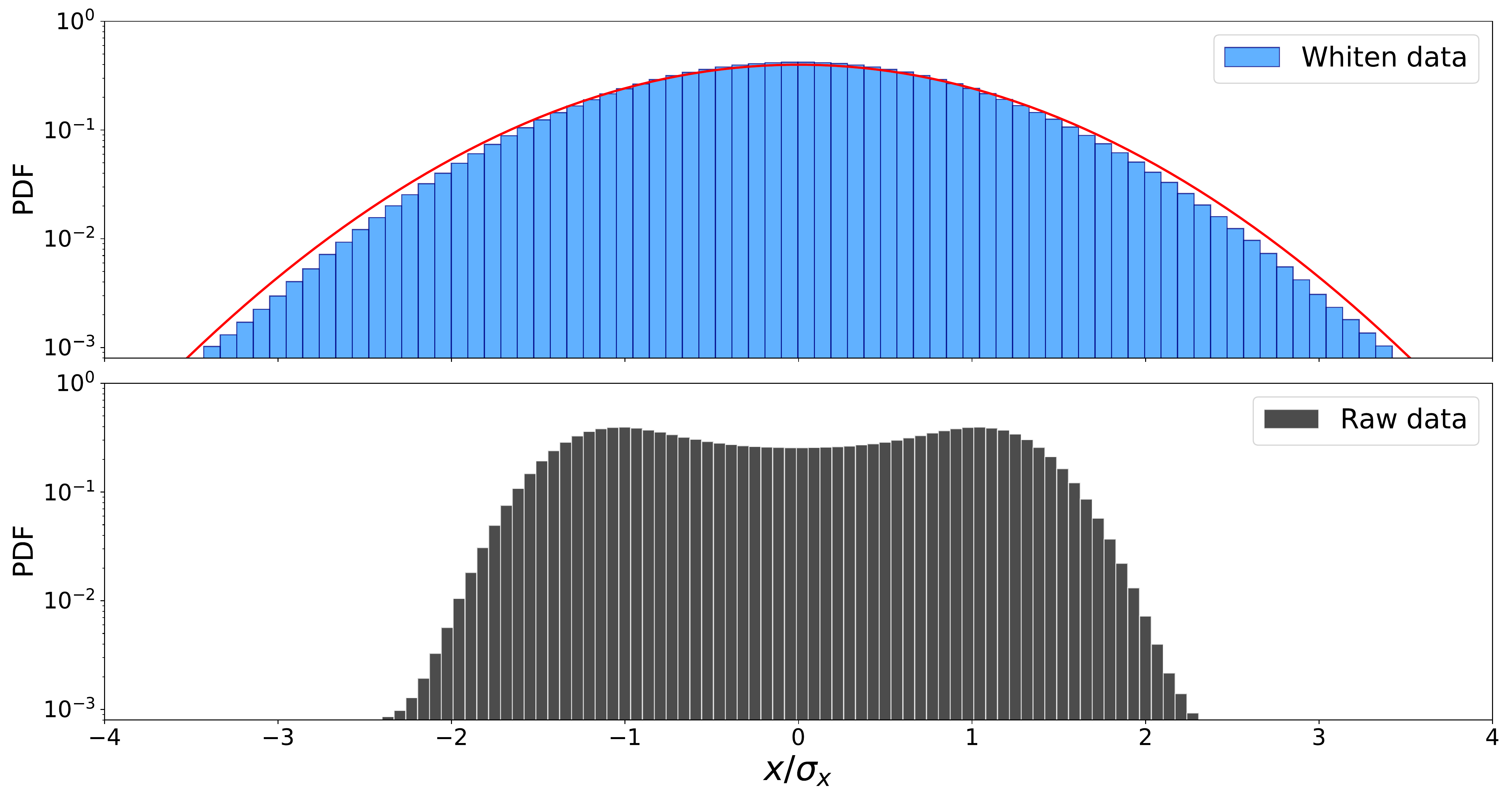}
\caption{ {\bf Probability distribution functions of normalized background and its whitened noises before the event 29 July 2017.}
In lower panel, probability distribution functions (PDF) of normalized background noise $x(t) \equiv x(t)/\sigma_x$ estimated from $\sim 3\times 10^4$ data points is plotted in log-linear scale. In upper panel, the PDF of its whitened time series is depicted which has a Gaussian. }
\label{F55}
\end{figure}

In summary, our aim in this work has been to introduce a new approach for the extraction of waveforms from GW events. Moreover, we estimate the physical time delays between the arrivals of gravitational waves at the GW detectors. 
The standard whitening procedure (in window size of $8 ~sec$) allows us to use the standard methods of false-alert rate estimation (see \cite{Hanna2008} and the references therein). The signal-to-noise ratio distribution is shown to be a Rayleigh distribution for the whitened data; see figure (\ref{F55}). We also provide the cross-correlation coefficients between the extracted waveforms in figure \ref{F33} and the simulated waveforms generated by {\it PyCBC} library for the events in the first, second, and third observing runs of O1, O2, and O3a, given in figure \ref{F5} \cite{pycbc}. Moreover, the absolute value of the cross-correlation coefficients (R) of the extracted waveforms for {\it H and L} detectors are presented in Table 1.  


The physical time delay between the amplitudes of the two detectors' signals is a complicated function of the orientation of the source, the orbital plane of the binary system, and the antenna patterns of the detectors. However, the Hilbert-Huang transform and Hilbert spectrum provide the instantaneous frequency of the extracted waveform independent of these geometric details.
 Therefore we are calling the estimated time delay as ``physical time delay".
In Methods, we generalized our approach to bivariate time series in which the extracted waveforms are estimated using the information provided by other detectors. We employed our generalized approach in the extraction of all waveforms in this study (Methods).
Although, we demonstrate the efficiency and applicability of our method in GW detection, however, this approach has the potential for extraction of event waveforms in many fields of science,
ranging from neuroscience, physics to biology.

\newpage
\section*{Materials and Methods}

\subsection*{Contemporary approaches to extract GWs}
Contemporary approaches to extract GWs can be divided into two major categories. 

{ \it Template-dependant methods}: This seeks to find GW events from the statistical comparison between LHV datasets and a range of simulated GW templates in time or frequency domains. The most prominent candidate method in this regard is using match filtering technique that tries to match the time-series from detectors to a wide range of pre-generated GW template waveforms, under certain physically meaningful criteria, through an optimal match filter design \cite{sofa1}. This technique is well developed to search for a GW event included in the non-stationary stochastic background noise.
Another technique introduced in Ref. \cite{sofa4} is to first find the GW events and then tries to extract a clean coalescence signal from noisy data.  It is also accompanied with some presumptions related to a common gravitational waveform behavior. There is also another approach based on cross-correlation of detectors in the time domain \cite{sofa3}. In this work, a simple measure consisting of the cross-correlation of detectors in short time intervals is introduced. The problem regarding this technique is that in contrast to the matched filter approach which is mainly based on cross-correlation in the frequency domain, here the correlation is measured in the time domain, thus demanding extra care. It means that the techniques used to improve the SNR should be carried out carefully to give rise to the possibly existing signal while suppressing the noise. This task alone, as is discussed and demonstrated in our work, needs to be implemented with much care since it is very sensitive to the true estimation of the non-stationary, non-Gaussian noise statistics. And also considering the abundance of instrumental glitches in the detectors with similar structures, the chances of detecting a glitch as a transient can be considerable. Possible correlations of the noise between detectors can be misleading too.

{\it Template-free methods}: On the other hand these methods try to find the presence of GW events without any prior assumption about GW templates. In the wavelet reconstruction method, although it is true that wavelet reconstruction is able to obtain clean GW event-waveforms, it highly depends on the selection of the appropriate mother wavelet function which could limit the attempts for general noise suppression problems in real-world \cite{sofa2}


The advantage of our proposed method compared to other common algorithms in the field is that our method can detect clean shape GW merger events without any prior information about the presence of the event. We start with the estimation of the noise variance in the Fourier domain from the signal after common pre-possessing steps. Then it updates the noise variance using an add-overlap method and tries to remove the noise based on this estimation. We also generalized this approach to calculate the noise statistics using a network of arrays (i.e. by updating the noise variance with a conditional joint probability bayesian estimator described in Eqs. 9-11, see below). Although this method works well to detect binary black hole mergers, we could not detect reliable results for binary neutron stars. The reason is due to the merger time duration in binary neutron stars that raise a challenge for the time constant that we consider for our add-overlap step.  We showed that the results are consistent with events reported by LIGO (see cross-correlation coefficients presented in figure 5). 
Moreover, due to the stochastic characteristic of the background noise detected in different detectors and the different angles of arrival of events in detectors, we estimate time delays between detectors in the frequency domain using the Hilbert-Huang transform. This helps us find correct physical time delays for each event compared to time delays estimated from cross-correlation coefficients in the time domain. After deriving of instantaneous frequencies using the Hilbert-Huang transform, we are looking for chirp waveforms where the power increases monotonically in frequency and time. 

 At first we split the signal into 8s windows to apply prepossessing steps. Then, we apply an add-overlap method on 0.06s windows in the Fourier domain to estimate expected values of the noise during the noise estimation step. Finally, we reconstruct the signal in the Fourier domain from summation over all 0.06s windows and estimate the Wiener gain function for the larger windows (8-sec windows) to extract the noise from the raw data.  

We did not obtain a clean extracted signal for binary neutron star mergers. Thus, we pick the frequency band $30-500 Hz$ for our estimation to focus on binary black hole mergers.

\subsection*{Details of the event-waveform detection method.}
In this section, we explain in detail the method introduced in the main body of the manuscript.

Our event-waveform detection approach is based on the statistical comparison in the time-frequency domain of background noise and a segment of the data that may includes the GW event. The steps are:

(i) First, we use a high-pass filter to filter out frequencies below $30 \,\mathrm{Hz}$. We also use notch filtering to eliminate certain high amplitude spectral lines, potentially disrupting the attempt to search for GW event-waveforms \cite{abbott2020guide}.


(ii) We whiten the raw data that is a technique commonly used in astrophysics and cosmology. Whitening is equivalent to flattening the spectrum of a given signal, allowing all the bands to participate equally in the power spectrum of a given time series. We applied whitening over time windows of length $8$ {\it sec}.


(iii) Finally, we employ a template-independent method to suppress the background noise. This includes our generalized implementation of a two-step decision-directed noise reduction method proposed in \cite{IMCRA} and an iterative Wiener filtering as the gain function and a noise estimation method based on recursive averaging algorithms considering GW events presence uncertainty \cite{W2-2,IMCRA,W2,W3}. 

For convenience, we first explain the decision-directed noise reduction method and the $a\ priori$ SNR used in Wiener filtering gain function described in Ref \cite{W2}. Then we introduce the noise power spectrum density (PSD) estimation method used in our noise reduction algorithm.

Let $x(t)=y(t)+n(t)$ be the noisy time series that we obtain from pre-processing steps, which consists the GW waveform denoted by $y(t)$ and the stochastic background noise denoted by $n(t)$. We assume that $y(t)$ and $n(t)$ are statistically independent, and both are zero mean time series. It is clear that in practice, we only have access to the noisy time series $x(t)$ collected during LHV observations, and the presence of a GW event $y(t)$ is unknown. However, we can estimate the noise variance during time frames in which no GW events are reported. We note that the nature of noise in advanced LHV time series is highly non-stationary and has a nonuniform effect so that different frequency bands have effectively different SNRs \cite{abbott2020guide}. As a result, in order to be able to conduct a statistical comparison between consecutive time frames in each frequency bands, we continue our analysis in time-frequency domain using short-time Fourier transform (STFT) over windowed data segments as

\begin{equation}
X_{k}(i,j)=Y_{k}(i,j)+N_{k}(i,j)
\label{eq:2}
\end{equation}
where index $k$ indicates the corresponded detector in LIGO-VIRGO network array, $X(i,j)$, $Y(i,j)$, and $N(i,j)$ represent the $i-{th}$ time frame and $j-{th}$ spectral components over time-frequency domain of the noisy time series, GW event-waveform, and background noise, respectively. Owing to the assumed independence of the GW waveform and the stochastic background noise, the periodogram of the noisy time series is approximately equal to the sum of the periodograms of GW signal and noise, respectively, that is:
\begin{equation}
|X_{k}(i,j)|^2\approx|Y_{k}(i,j)|^2+|N_{k}(i,j)|^2 \quad .
\label{eq:5}
\end{equation}
Once we obtain the magnitude and the spectrum of the desired GW waveform $Y(i,j)$, we are able to reconstruct the final GW event-waveform in the time domain $y(t)$, using inverse fast Fourier transform (iFFT). In practice, no direct solution for the spectral estimation exists. As the result, most common signal enhancement techniques require the evaluation of two parameters as the {\it a posteriori} and the {\it a priori} SNRs for the noisy components and GWs spectral components, respectively. Then, an estimate of $Y(i,j)$ is subsequently obtained by applying a spectral gain function as:
\begin{equation}
\hat{Y}_{k}(i,j)=G{\big(}\xi_{k}(i,j), \gamma_{k}(i,j),q_{k}(i,j){\big)} ~ X_{k}(i,j)
\label{eq:3}
\end{equation}
where $\xi(i,j)$ and $\gamma(i,j)$ are known to be $a\ priori$ and $a\ posteriori$ SNRs in $i-{th}$ time frame and $j-{th}$ frequency band, respectively, defined as:
\begin{equation}
\xi_{i,j} \coloneqq \frac{E[|Y_{i,j}|^2]}{E[|N_{i,j}|^2]} ; \qquad
\gamma_{i,j} \coloneqq \frac{|X_{i,j}|^2}{E[|N_{i,j}|^2]} \quad.
\label{eq:4}
\end{equation}
Here $E[.]$ is the expectation value operator and $q_k$ shows the signal absence probability of the time series related to $k-{th}$ detector in time-frequency domain, which is set to be zero here, meaning that we assume the existence of signal in all short time-frequency segments that we are analyzing. The function $G$ is the spectral gain function that should be applied to each short-time frame of the spectral component of the signal to obtain the spectral component of the clean signal. The choice of the gain function determines the gain behavior that determines the level of noise reduction in this method. We applied a Wiener gain function as described in Eq. (8) see below, and Ref. [6].  

We note that, one can't estimate Eq. (3) directly because we don't know the value of $Y_k(i,j)$. However, we can estimate the a-priori SNR and $Y_k(i,j)$ respectively using the assumption that a-priori SNR for frame $i$ can be estimated from Eq. (7) that is a decision-directed step (as known as first-order recursive function) to update $\xi{(i,j)}$ from previous frame $i-1$.

Substituting Eq. 10 into Eq. 9 yields:
\begin{equation}
\xi(i,j) =\frac{|X(i,j)|^2-E[|N(i,j)|^2]}{E[|N(i,j)|^2]}=\gamma(i,j)-1\quad .
\label{eq:6}
\end{equation}
By tracking the noise power spectrum density (PSD), we are able to estimate the $a\ priori$ and $a\ posteriori$ SNRs. In order to be able to bridge the broadest peak in the GW signal, we choose relatively wide enough window length -$0.06 sec$- for the estimation. However, the periodogram $|X(i,j)|^2$ of the noisy time series fluctuates very rapidly over time. As the result, we use a first-order recursive version of the $\gamma(i,j)$ estimator called decision-directed approach. Given that the $a\ priori$ and the $a\ posteriori$ SNRs can be estimated as:

\begin{equation}
\hat{\gamma}(i,j)=\frac{|X(i,j)|^2}{E[|N(i,j)|^2]}
\label{eq:7}
\end{equation}
and
\begin{equation}
\hat{\xi}(i,j)=\lambda\frac{|\hat{Y}(i-1,j)|^2}{E[|N(i,j)|^2]}+(1-\lambda)P[\hat{\gamma}(i,j)-1]
\label{eq:8}
\end{equation}
where $P[.]$ denotes the half-wave rectification and $\hat{Y}(i-1,j)$ is the estimated waveform spectrum at previous time frame in each frequency bands. We used half-wave rectification during the {\it a priori} and the {\it a posteriori} estimation using harmonic regeneration noise reduction -{\it HRNR}- method  as proposed in Ref. \cite{hrnr}. Using {\it HRNR} to estimator SNRs has shown to be more accurate in reducing background noise while it helped us to prevent the distortion \cite{hrnr}. The control parameter $0<\lambda<1$ is the weighting factor in decision-directed approach which determines the smoothness of the estimation. 
 $\lambda$ also is a smoothing factor that determines the smoothness of the a-priori estimator. Higher values of $\lambda$ give smoother results. We choose $\lambda = 0.95$ as a trade-off between noise reduction and the event distortion.

Then, using the Eq.\ref{eq:3}, we are able to obtain the estimated clean GW waveform $\hat{Y}(i,j)$. The gain function we used in this paper is based on Wiener filtering as \cite{W2}:
\begin{equation}
G(\xi_{i,j}, \gamma_{i.j})=\frac{\hat{\xi}_{i,j}}{1+\hat{\gamma}_{i,j}} \quad .
\label{eq:9}
\end{equation}
So far, we assumed that an estimation of the noise spectrum, $N(i,j)$ is available and the noise PSD is given by $\sigma_{n}^2=E[|N(i,j)|^2]$. However, in general, we only have access to the noisy time series which yields having only the noisy signal spectrum $X(i,j)$ to be known, while the PSDs of the GW waveform $|Y(i,j)|^2$ and the background noise $\sigma_{n}^{2}$ are still to be unknown. Thus, both $a\ posteriori$ SNR and $a\ priori$ SNRs have to be estimated. In literature, there are different approaches to estimate the noise PSD $\sigma^{2}$ \cite{W2-2}.

(i) Minimum statistics methods, assume that the noisy time series's power in each frequency band often decays to the noise power level. Thus, one can estimate the noise
power spectrum by searching for minimum values in the noisy time series PSD, at each segment over the time-frequency domain.

(ii) Histogram based methods, assume that the most frequent energy levels in the power spectrum distribution at each segment over the time-frequency domain correspond to the noise level. As a result, the noise PSD is estimated by finding maximum values in the power spectrum distribution.

(iii) The time-recursive averaging methods are developed based on the assumption that the presence of noise has a nonuniform effect on the spectrum of a given time series so that different frequency bands have effectively different SNRs. Consequently, one can estimate the noise spectrum in each time-frequency segments whenever SNR is very low. This leads to a noise estimation approach based on a weighted average statistical comparison between the past estimates and the present noisy time series spectrum \cite{W2-2,IMCRA}. In this manner, the weighted average variable updates adaptively based on the noise and the GW event's absence or presence.

Here we use a generalized version of the Improved Minima Controlled Recursive Averaging (IMCRA) algorithm to obtain more accurate estimator due to the non-stationarity nature of the background noise \cite{IMCRA}. In general, the presence or absence of a GW event or equivalently the detection of such event in $j^{th}$ frequency bin, in time-frequency domain of the data obtained from $k^{th}$ detector in LVH network array, is cast as a detection problem which yields in two following hypotheses:

\begin{eqnarray}
\begin{aligned}
H_{0}^{j}(i,j): && GW_{ event~ absence}: \hskip 0.5cm X_{k}(i,j)&=&N_{k}(i,j) \cr \nonumber
H_{1}^{j}(i,j): && GW_{ event~ presence}: \hskip 0.5cm X_{k}(i,j)&=&Y_{k}(i,j)+N_{k}(i,j) \quad .
\end{aligned}
\label{eq:10}
\end{eqnarray}
The noise PSD in $k^{th}$ detector can be estimated in terms of minimum mean square error as:
\begin{eqnarray}
\hat{\sigma}_{n,k}^{2}(i,j) &=& E{|N_{k}(i,j)|^{2}}=E[\sigma_{n,k}^{2}(i,j)|X_{k}(i,j)] \cr \nonumber \\
&=&E[\sigma_{n,k}^{2}(i,j)|H_{0}^{j}]P(H_{0}^{j}|X_{k}(i,j))+E[\sigma_{n,k}^{2}(i,j)|H_{1}^{j}]P(H_{1}^{j}|X_{k}(i,j))
\label{eq:11}
\end{eqnarray}
where $P(H_{0}^{j}|X_{k}(i,j))$ denotes the conditional probability of GW event being absent in time-frequency segment $i,j$ given the noisy time series spectrum $X_{k}(i,j)$ collected in LHV network. Similarly, $P(H_{1}^{j}|X_{k}(i,j))$ denotes the conditional probability of GW event being present in time-frequency segment $i,j$ given the noisy time series spectrum $X_{k}(i,j)$ collected in LHV network. We can compute our conditional probabilities in Eq.\ref{eq:11}, using Bayesian rules as follows:
\begin{eqnarray}
P(H_{0}^{j}|X_{k}(i,j))&=&\frac{P(X_{k}(i,j)|H_{0}^{j})P(H_{0}^{j})}{P(X_{k}(i,j)|H_{0}^{j})P(H_{0}^{j})+P(X_{k}(i,j)|H_{1}^{j})P(H_{1}^{j})} \cr \nonumber
&\equiv& \frac{1}{1+r_{k}\Lambda_{k}(i,j)}
\label{eq:12}
\end{eqnarray}
where $r_{k}\equiv\frac{P(H_{1}^{j})}{P(H_{0}^{j})}$ is the ratio of the $a\ priori$ probabilities of GW event absence or presence, and $\Lambda_{k}(i,j)\equiv\frac{P(X_{k}(i,j)|H_{1}^{j})}{P(X_{k}(i,j)|H_{0}^{j})}$ is the likelihood ratio. Similarly for $P(H_{1}^{j}|X_{k}(i,j))$ we have:
\begin{eqnarray}
P(H_{1}^{j}|X_{k}(i,j))&\equiv& \frac{r_{k}\Lambda_{k}(i,j)}{1+r_{k}\Lambda_{k}(i,j)}
\label{eq:13}
\end{eqnarray}
Substituting \ref{eq:12} and \ref{eq:13} into \ref{eq:11} reads:
\begin{eqnarray}
\hat{\sigma}_{n,k}^{2}(i,j) &=& \frac{1}{1+r_{k}\Lambda_{k}(i,j)}E[\sigma_{n,k}^{2}(i,j)|H_{0}^{j}]+\frac{r_{k}\Lambda_{k}(i,j)}{1+r_{k}\Lambda_{k}(i,j)}E[\sigma_{n,k}^{2}(i,j)|H_{1}^{j}] \quad .
\label{eq:14}
\end{eqnarray}

When GW event is absent in frequency bin $j$ in $k^{th}$ detector, the term $E[\sigma_{n,k}^{2}(i,j)|H_{0}^{j}]$ is approximately equals to the PSD of the noisy time series, $|X_{k}(i,j)|^{2}$. Alternatively, we can approximate $E[\sigma_{n,k}^{2}(i,j)|H_{1}^{j}]$ using the PSD of the noise in previous frames, $\hat{\sigma}_{n,k}^{2}(i-1,j)$ \cite{W2-2,IMCRA}. Substituting these approximations into \ref{eq:14} gives the PSD estimate of the noise as:
\begin{eqnarray}
\hat{\sigma}_{n,k}^{2}(i,j) &=& \frac{1}{1+r_{k}\Lambda_{k}(i,j)}|X_{k}(i,j)|^{2}+\frac{r_{k}\Lambda_{k}(i,j)}{1+r_{k}\Lambda_{k}(i,j)}\hat{\sigma}_{n,k}^{2}(i-1,j) \cr \nonumber \\
&=& (1-\alpha_{k}(i,j))|X_{k}(i,j)|^{2}+\alpha_{k}(i,j)\hat{\sigma}_{n,k}^{2}(i-1,j) \quad .
\label{eq:15}
\end{eqnarray}
Eq. \ref{eq:15} has the form of a time-recursive noise estimator where $\alpha_{k}(i,j)= \frac{r_{k}\Lambda_{k}(i,j)}{1+r_{k}\Lambda_{k}(i,j)}$ is the smoothing factor \cite{W2-2}. Hence, under Gaussian assumption, $\Lambda_{k}(i,j)$ can be computed as:

\begin{eqnarray}
\Lambda_{k}(i,j)=\frac{1}{1+\xi_{k}(i,j)}exp\left(\frac{\xi_{k}(i,j)}{1+\xi_{k}(i,j)}\gamma_{k}(i,j)\right)
\label{eq:16}
\end{eqnarray}
where $\xi_{k}(i,j)$ and $\gamma_{k}(i,j)$ are the $a\ priori$ and the $a\ posteriori$ that for $k^{th}$ detector can be estimated using decision-directed estimator given by Eq. \ref{eq:7} and \ref{eq:8} \cite{W2, W2-2}. 

As we obtain spectral components of the desired GW waveform $Y$, we can reconstruct the denoised GW event-waveform in the time domain $\tilde{y}(t)$ using iFFT. We find that due to the non-stationarity of the time series under study, the noise frame of $\sim 10$ seconds duration adjacent to the mainframe in which GW events received provide good results for the average weighted smoothing factor. This leads to an acceptable estimation of the averaged noise spectrum and also avoid the spectral artifacts that may appear locally in time. We also demonstrate the data in the unit of its standard deviations of a normalized window comprising the processed noise and the processed GW event's time windows, which in a way can be considered as an illustration of the signal contribution in noisy signal.

In summary, our method is based on the assumption that the power of the noisy signal in each frequency band decays to the power level of the noise as described in Refs. [6-8]. Therefore, we can estimate the noise level in each frequency band by tacking the minimum of the power in the noisy signal. This method was first developed to estimate the presence of speech signals under noisy conditions. Here, we use this idea and try to generalize it to obtain the clean GW event waveforms without having any prior information about the existence of the GW event in the analysed data. Moreover, we use the Wiener gain function as described in Ref. [6] and combine it with a noise estimator algorithm described in Ref. [55] to obtain better results for the power spectrum of the noise during our estimation. Our steps after pre-possessing include noise estimation as described in Eqs.  (9-13). Then, we substitute the estimated expected value of $|N(i,j)|^2$ from this step into Eq. (7) and estimate the apriori SNR for $i^{th}$ time frame and $j^{th}$ frequency component. Finally, we substitute $\xi{(i,j)}$ into Eq. (8) which is the Wiener gain function and estimate the extracted signal.

\subsection*{Interpretation of parameter $\alpha$} 
 The total strain indicated by $x(t) = n(t) + \alpha y(t)$, the first term $n(t)$ represents the noise of detector and $y(t)$ is the signal. Since 
gravitational wave follows the wave equation as $\Box h_{\mu\nu} =0$, the solution of this equation depends on the 
distance from the source (in Minkowski Space) as $h_{\mu\nu} \propto 1/r$. So the coefficient $\alpha$ depends to $r$ as $\alpha\propto 1/r$. Now, we can write the time dependent $x(t)$ as 
$ x(t;r) = n(t) + (r_0/r) y_0(t)$
where $r$ is the distance of GW source and $r_0$ is a given distance from the observer and $y_0(t)$ is the signal of a GW source at 
the distance of $r_0$. In our analysis varying $\alpha$ within range of $10^{-3}$ to $5\times 10^{-3}$ means that we allow the distance of the 
GW source to change by fifty times. In other words,  $\alpha_1/\alpha_2 = 50$ then $r_2/r_1 = 1/50$. For instance, for the event "GW150914" located at distance of $r_0 = 800 Mpc$, the time series at any arbitrary distance of $r$ is related 
to $\alpha$ as $r = 800 Mpc/\alpha$. For the maximum peak of strain of $4\times 10^{-20}$ in the template, using $\alpha = 0.005$ means that distance 
of the source is located at few Gpc (taking into account the cosmological calculation) with the strain of $2\times 10^{-22}$. \\

\subsection*{Cross correlation coefficients of extracted waveforms with those inferred by LIGO} 
To check the similarity of our extracted waveforms with those inferred by LIGO, we have estimated the Pearson correlation coefficients of the waveforms with the best fits among the waveforms simulated based on posterior BBH parameters reported by LIGO \cite{De2019, abbott2021gwtc, PRX, venumadhav2019new}. To this end, we have used the PyCBC package \cite{pycbc} to generate the time domain waveforms for each detector, using the posterior source files reported by LIGO for each event, including the parameters: primary and secondary masses, cartesian spin elements for each object, distance, inclination, declination, right ascension and polarization. The approximates used here are "IMRPhenomPv2" for all events except for GW190814 and GW190521, for which "IMRPhenomD" and "IMRPhenomPv3HM" is used, respectively. We note that by best fit, we are pointing to the one simulated waveform that returns the largest correlation coefficient with our extracted waveforms, which is chosen among thousands of posterior sets, typically. In Fig. \ref{F5}, the absolute values of the correlation coefficient between all extracted waveforms and their associated simulated waveforms that match the extractions the best, are reported for both (L and H) detectors. We note that the correlation coefficients are evaluated in the $0.2s$ time-windows.

\subsection*{ The $\chi^2$ time-frequency discriminator }

In this paper signal to noise ratio (SNR) is defined as the ratio of extracted waveform and the mean value of processed background noise, $\rho(i,j) = |\hat{y}(i,j)|^2/E[|\hat{n}(i,j)|^2]$, where the frequency band is divided in p-bins. We employ $\chi^2$ time-frequency discriminator for gravitational wave detection as follows. In each frequency bin the normalized SNR is $z_i$ and we define $z=\sum_{i=1} ^p z_i$ and $\chi_0^2 = p \sum_{i=1} ^p (z_i - \langle z \rangle)^2$. 
Provided that the SNRs in each bin have Gaussian distribution, it is shown that $P_{\chi^2 < \chi_0^2 } = \frac{\gamma(p/2-1/2, \chi_0^2/2)}{\Gamma(p/2-1/2)}$, where $\gamma(\cdots)$ is the incomplete gamma function \cite{chi2}. High and low values of  $P_{\chi^2 < \chi_0^2 }$ are indicators of chirp and spurious signals. Intuitively, this stems from the fact that in chirp signal in the chosen frequency band, all SNRs have finite values and their dispersion are small, however in spurious signal, some bins have high SNR and in other bins have smaller, then there is the high value of dispersion.  

To check the Gaussianity of SNRs in each frequency bin, we estimate the SNR for a random set of 560 captures from Hanford detectors during the O2 (from GPS Time 1164556817  to GPS Time 1187733618) and O3a (from GPS Time 1238166018 to GPS Time 1253977218) run
\cite{abbott2021gwtc, PRX}. The duration of each capture was 4096 sec and data points were sampled at $4$KHz. The total time duration was $637$h. A list of the name of time series can be found at Supp. Mat. To estimate SNRs, we first applied a highpass filter with a frequency pass $> 30$ Hz. We assumed the first 10 sec of each time series as the initial noise to start our noise reduction algorithm. Finally, we estimated SNRs of each time series in $p=16$ frequency bins between [30 - 512 Hz]. 

\begin{figure}[H]
\includegraphics[width=\textwidth]{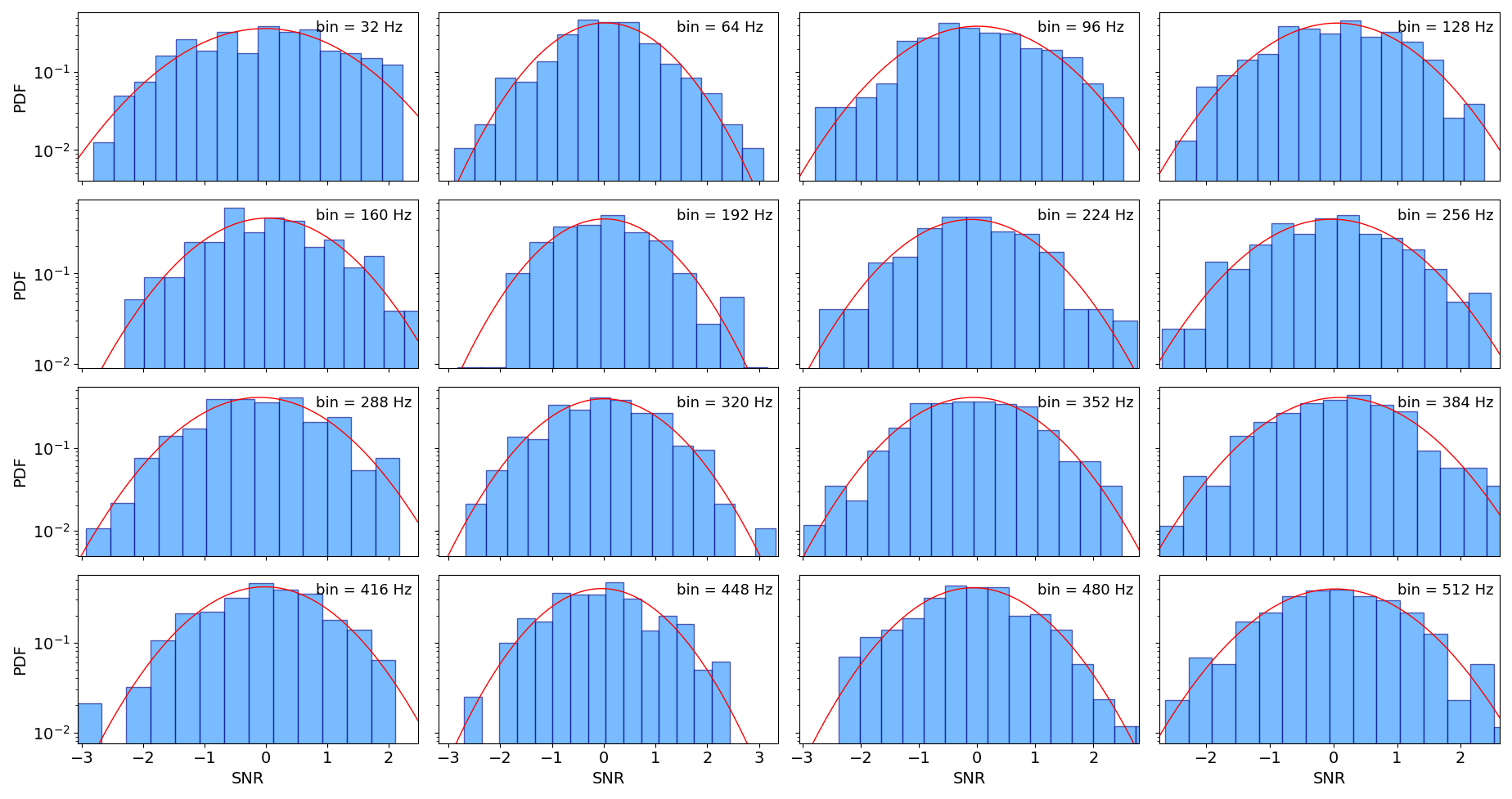}
\caption{ The results show  PDF of normalized SNRs for each frequency bin. The total time duration of analysed time series is about $637$ h. In each subgraph a Gaussian PDF with unit variance is plotted for comparison. }
\label{FR}
\end{figure}

We find that our defined SNR in each bin follows from the Gaussian distribution, which is needed to employ the $\chi^2$ time-frequency discriminator test (see Fig.\ref{FR}). 

\subsection*{Time delays between the arrivals of gravitational waves to the detectors.}

\vskip 0.5cm


We provide the value for the time delays between the arrival of gravitational waves to the detectors from analyzing the processed data. We use the Hilbert spectrum for the physically meaningful time-lag for the two detectors, i.e., $<10\,\mathrm{ms}$ for instance, for L and H. 


\subsubsection*{Hilbert Transform and Hilbert spectrum}

Let us consider extracted waveforms in LIGO and Virgo detectors $x(t) \coloneqq \tilde{y}(t)$ and determine their local phases using for instance Hilbert transform or marked events method \cite{Tabar2019}.
To determine the local "phase" of time series, we apply the Hilbert transform to process $x$
as
\[
y(t) = \frac{1}{\pi} P \int _{-\infty} ^{\infty} \frac{x(t')}{t-t'} dt'
\]
where $P$ is the Cauchy principal value of the integral. We define analytical signal $z(t) = x(t) + i y(t) = a(t) e^{i \phi(t)}$, where $a(t) = [ x(t)^2 + y(t)^2]^{1/2}$ and therefore local phase is given by $\phi(t) = \tan^{-1} (y(t)/x(t))$. From the local phase one can calculate the local frequency $f(t)$ via its time derivative $f(t)=\frac{1}{2\pi}\frac{d}{dt} \phi(t)$.

\begin{figure}[H]
\includegraphics[width=.6\textwidth]{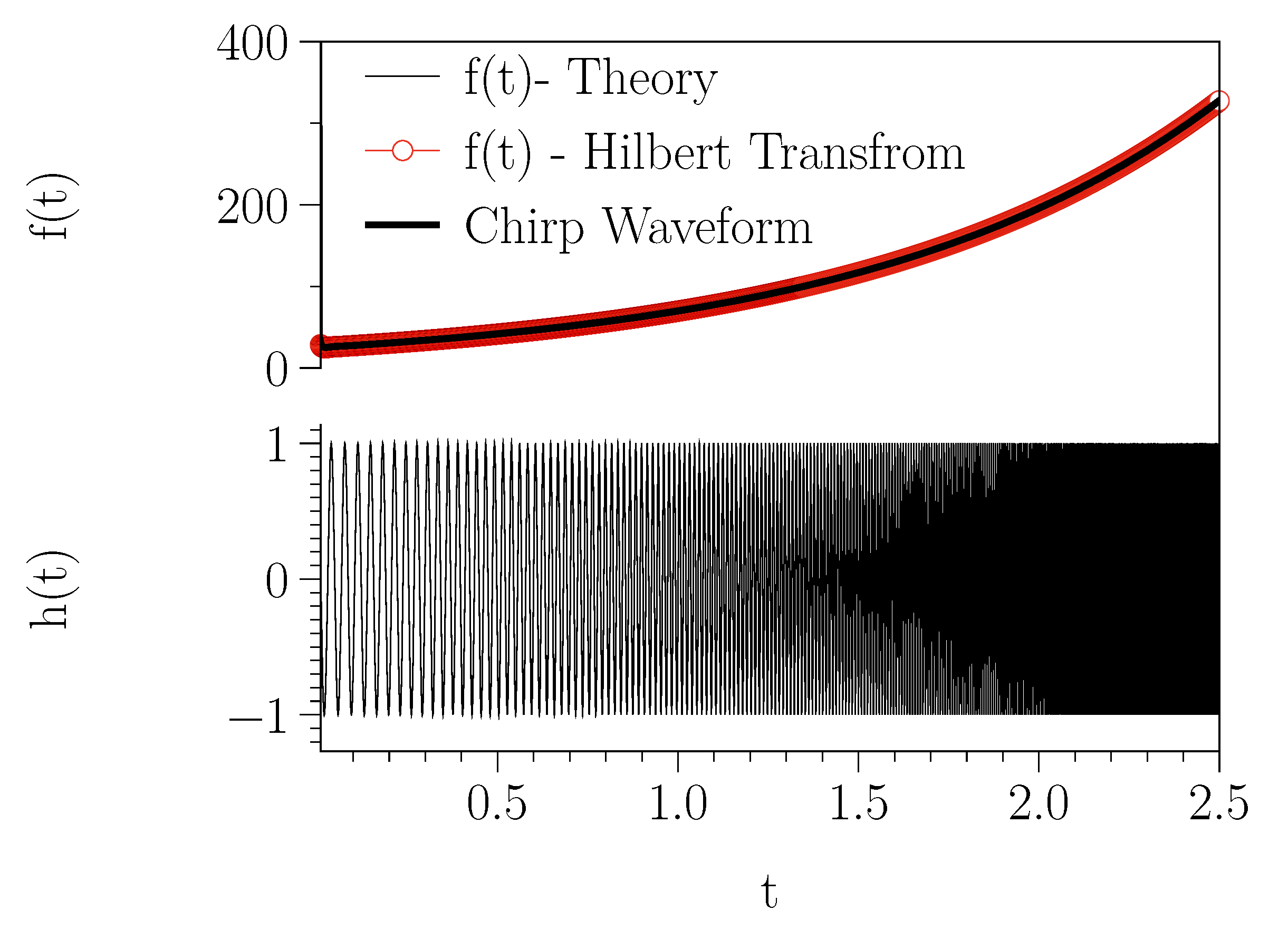}
\caption{ {\bf A chirp waveform and its Hilbert transform.}
A sinusoidal exponential chirp waveform and its theoretical time-dependent frequency as well as estimated frequency from Hilbert transform of discretized waveform. The waveform is plotted for the parameters are $k=(f_0/f_f)^{1/T}=2.8$, with $f_0 =25$, $f_f =550$ and $\phi_0=0$. }
\label{F7}
\end{figure}

 As an example, consider a chirp waveform as
\[
x(t) \coloneqq h(t) = \sin\left[\phi_0 + 2\pi f_0 \left(\frac{k^{t} - 1}{\ln(k)}\right) \right],
\]
where the time dependent frequency is given by $f(t) = f_0~ k^t$, $\phi_0$ is the initial phase, $f_{0}$ is the starting frequency (at t=0), and $k$ is the rate of exponential change in frequency. In figure (\ref{F7}), we plot the waveform as well as the time dependent theoretical frequency and also the one estimated from Hilbert transform of discretized with $dt=10^{-5}$. The waveform is plotted for the parameters as $k=2.8$, $f_0 =25$ and $\phi_0=0$. We repeat the Hilbert transformation for initial phase of $\phi_0=\pi/3$, and notice that the time-dependent frequency obtained via Hilbert transformations with $\phi_0=0$ and $\phi_0=\pi/3$ are the same.


\begin{figure}[H]
\includegraphics[width=.7\textwidth]{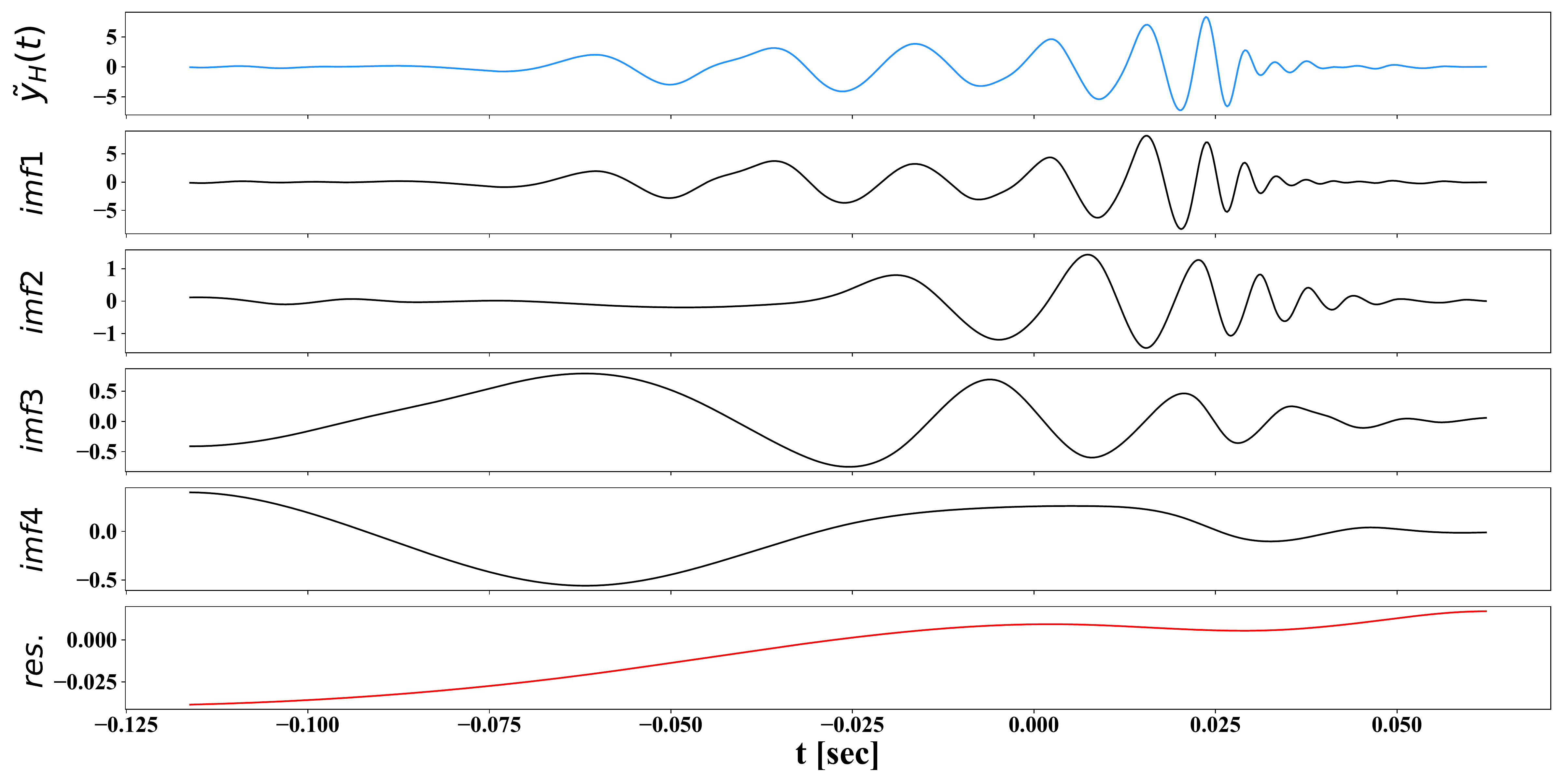}
\caption{ {\bf Intrinsic mode functions of extracted waveform of event GW150914 in H.}
The intrinsic mode functions (IMF1 to IMF4) and residual (res.), for the extracted waveform $\tilde{y}(t)$ of the event GW150914 in H from empirical mode decomposition. 
}
\label{F8}
\end{figure}

To estimate frequency-time relationship for the extracted waveform of $\tilde{y}(t)$, we can apply Empirical Mode Decomposition (EMD) (see [49] \cite{EMD} for details). The aim is to decompose the original signal into a hierarchy of Intrinsic Mode Functions (IMF) that separate the
signal's different frequency components by counting the maxima and zero crossings. In figure (\ref{F8}), we plot
the IMFs of the extracted waveform $\tilde{y}(t)$ for the event GW150914 in H-detector, using the EMD method.



To estimate local frequency or frequency-time relationship, we then apply Hilbert transform to each IMF  components of the extracted waveform $\tilde{y}(t)$, excluding the final residuals, where one can express it as the real part ($\Re$) of the sum of the Hilbert transform of all the IMF components \cite{EMD},
\[
\tilde{y}(t) = \Re \sum_{j=1} ^N a_j(t) \exp\{i \int^t \omega_j(t') dt'\}
\]
where $a_j(t)$ and $\omega_j(t)$ are the amplitude and the frequency of the jth IMF component, respectively. Thus, the amplitude is a function of time and frequency. The frequency-time relationship of the amplitude is known as the Hilbert spectrum. The inherent characteristics of a nonlinear and/or non-stationary waveform can be identified from the Hilbert spectrum. Local frequencies of all extracted waveforms in this work are estimated from the Hilbert spectrum. One can verify that the Hilbert spectrum and the local time-frequency, such as a sinusoidal, exponential chirp waveform, provide the same frequency-time relationship with different initial phases $\phi_0$.







\bibliography{GW-Final}

\section*{Acknowledgements}

We want to thank Bahram Mashhoon for his insightful comments and discussions, which helps us improve this work's presentation in ideas and clarity. We are grateful to Abolhassan Vaezi, Mehdi Torabian, Jahed Abedi, and Hassan Khalvati for useful discussions and would like to thank Ali Rezakhani for his critical reading of the manuscript. \\
This research has made use of data, and web tools obtained from the Gravitational Wave Open Science Center (https://www.gw-openscience.org/ ), a service of LIGO Laboratory, the LIGO Scientific Collaboration and the Virgo Collaboration. LIGO Laboratory and Advanced LIGO are funded by the United States National Science Foundation (NSF) as well as the Science and Technology Facilities Council (STFC) of the United Kingdom, the Max-Planck-Society (MPS), and the State of Niedersachsen/Germany for support of the construction of Advanced LIGO and construction and operation of the GEO600 detector. Additional support for Advanced LIGO was provided by the Australian Research Council. Virgo is funded, through the European Gravitational Observatory (EGO), by the French Centre National de Recherche Scientifique (CNRS), the Italian Istituto Nazionale di Fisica Nucleare (INFN) and the Dutch Nikhef, with contributions by institutions from Belgium, Germany, Greece, Hungary, Ireland, Japan, Monaco, Poland, Portugal, Spain.\\
We also would like to thank two anonymous referees for providing thoughtful comments that helped us to improve our results. \\ 
This paper is dedicated to the memory of Professor Jalal Samimi (1941-2020) whose innovative and groundbreaking early works in high energy astrophysics and Gamma-ray astronomy resulting in identifying one of the first black hole candidate in the Milky Way Galaxy, GX339-4, \cite{SAMIMI1979} has inspired many researchers in the field of astrophysics. 

\section*{Funding}
S.B. acknowledges partial support by Sharif University of Technology's Office of Vice President for Research and Technology under Grant No. G960202.

\section*{Author contributions statement}
The problem was conceived by H. A., S. R., S. B., and M. R. R. T. The computer codes were written by H. A., and A. A. Methods has been written by A. A. All authors analyzed the results and reviewed the manuscript.

\section*{Competing interests}
The authors declare no financial or non-financial competing interests.

\section*{Data availability}

All results are reported for public data of LIGO with sample-rate $4096 \mathrm{Hz}$. Data on extracted waveforms are available upon request.

\section*{ Code availability}
All codes developed in this study are available upon request.

\end{document}